\documentclass[twocolumn,amsmath,amssymb,10pt,floatfix,superscriptaddress,a4paper,letterpaper,fleqn,nofootinbib]{revtex4-1}
\usepackage{amssymb}
\usepackage{graphicx}
\usepackage{epsfig}
\usepackage{dcolumn}
\usepackage{array}
\usepackage{bm}
\usepackage{fancyhdr}
\usepackage{floatflt}
\usepackage{longtable}
\usepackage{threeparttable}
\usepackage{multirow}
\usepackage{float}
\usepackage{afterpage}
\usepackage{color}
\usepackage{textcomp}
\usepackage{multirow}
\usepackage{mathdesign}
\usepackage{booktabs}
\usepackage{paralist}
\usepackage{float}
\usepackage[caption = false]{subfig}

\newcommand{\nuc}[2]{$^{{\mathrm{#1}}}${#2}}

\usepackage[warnundef]{jabbrv}
\DefineJournalAbbreviation{Test}{Tes} 
\setlongtables
\definecolor{Darkgreen}{rgb}{0,0.4,0}
\usepackage[breaklinks=true,linkbordercolor={1 1 1}]{hyperref}
\hypersetup{
colorlinks = true,    
linktoc    = all,     
linkcolor  = blue,    
citecolor  = Darkgreen
}
\def\etal{\emph{et~al.}\ }

\usepackage{lipsum}

\begin{document}
\setcounter{page}{1}
\title{
 \qquad \\ \qquad \\ \qquad \\  \qquad \\ \qquad \\ \qquad \\
Evaluation of $^{86}$Kr Cross Sections For Use in Fusion Diagnostics
}

\author {M.~Vorabbi}
        \email[Primary author: ]{mvorabbi@bnl.gov}
  	\affiliation{National Nuclear Data Center, Brookhaven National Laboratory, Upton, NY 11973, USA}
\author {G.P.A.~Nobre}
  	\email[Corresponding author: ]{gnobre@bnl.gov}
  	\affiliation{National Nuclear Data Center, Brookhaven National Laboratory, Upton, NY 11973, USA}
\author {D.A.~Brown}
  	\affiliation{National Nuclear Data Center, Brookhaven National Laboratory, Upton, NY 11973, USA}
\author{A.M.~Lewis}
	\thanks{Current affiliation: Naval Nuclear Laboratory.}
	\affiliation{Rensselaer Polytechnic Institute, Gaerttner LINAC Center, 110 8th Street, Troy, New York 12180}
\author{E.~Rubino}
	\thanks{Current affiliation: National Superconducting Cyclotron Laboratory, Michigan State University.}
	\affiliation{Department of Physics, Florida Atlantic University, 777 Glades Road, Boca Raton, FL 33431, USA}
\author{S.~Mughabghab}
	\thanks{\textit{In memoriam}.}
  	\affiliation{National Nuclear Data Center, Brookhaven National Laboratory, Upton, NY 11973, USA}

\date{\today}

\begin{abstract}{
The National Ignition Facility at Lawrence Livermore National Laboratory uses $^{86}$Kr as a diagnostic tool to measure the neutron flux produced by fusion reactions.
As krypton is chemically inert, it can be implanted directly into the fuel capsule, and the reaction products can be measured to determine the flux of fusion neutrons.
$^{86}$Kr cross sections also provide model constraints for the $^{85}$Kr branching point in the s-process and  the neutron flux in stars.
In this work, experimental data on the neutron production, radiative capture, inelastic scattering, and total cross sections of $^{86}$Kr were used in conjunction with the fast region
nuclear reaction code EMPIRE and a new resonance-region evaluation to produce a new evaluation of neutron-induced reactions on $^{86}$Kr.
For the EMPIRE calculations, we fitted the optical model potential up to 12 MeV to simultaneously reproduce the experimental data for the total cross section and the main inelastic
gamma transition from the $2^+$ state to the $0^+$ ground state. This choice turned out to be very effective and produced a very good agreement of our calculations and the
experimental data also for all the other inelastic transitions. For energies above 12 MeV, due to large fluctuations and uncertainties in the total cross section data, we preferred to adopt the
Koning-Delaroche global spherical optical model potential. Always in this energy regime, the $(n,2n)$ channel is dominant and the physics is dictated by the pre-equilibrium.
In this energy regime we adopted the exciton model that was fitted to reproduce the available data in the $(n,2n)$ channel and consitently produced a good description of the
high-energy tail of the data for the inelastic gamma transitions.
With these models and corrections to the structure of $^{86}$Kr, the evaluated cross sections matched the experimental data.
The new evaluation has been submitted for incorporation in the next release of the ENDF/B nuclear reaction library.
}
\end{abstract}
\maketitle

\lhead{Evaluation of \nuc{86}{Kr} $\dots$}     
\chead{NUCLEAR DATA SHEETS}                  
\rhead{M. Vorabbi \textit{et al.}}        
\lfoot{}
\rfoot{}
\renewcommand{\footrulewidth}{0.4pt}



\section{Introduction}
\label{Sec:Introduction}



The $^{86}$Kr isotope is important in applications for both fusion technology and astrophysics.
As a chemically inert noble gas, krypton can be frozen and implanted in the fuel capsules used at the National Ignition Facility (NIF) at Lawrence Livermore National
Laboratory~\cite{NIFPaper}.
During a laser shot at NIF, the Kr nuclei are directly exposed to the neutron flux at the location of the burn, providing the most accurate picture of the fuel areal density.
The RAGS (radiochemical analysis of gaseous samples) detector is then used to collect the krypton gas and analyze it for certain decay products that are produced by neutron-induced
reactions with the fusion neutrons.
The $^{86}$Kr(n,2n)$^{85m}$Kr and $^{86}$Kr(n,$\gamma$)$^{87}$Kr reactions are of particular interest for this application, as the resultant nuclei of both reactions de-excite with
half-lives and energies that are convenient for measurement.
The RAGS detector allows for the determination of the number of each isotope produced in the shot, which can be used to calculate the flux of neutrons caused by the fuel burn if the
cross sections of the reactions are well understood.

\begin{figure}[t]
\includegraphics[width=1\columnwidth]{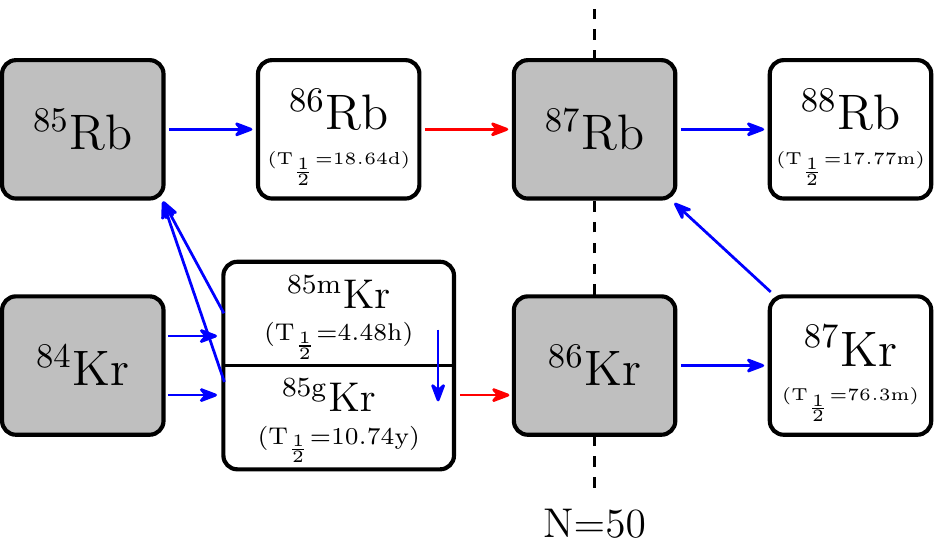}
\caption{Branching diagram of the s-process showing how $^{85}$Kr can either undergo beta decay to become $^{85}$Rb or neutron capture to become $^{86}$Kr, and how the
$^{86}$Kr(n,$\gamma$)$^{87}$Kr cross section is important for calculating the abundance of $^{87}$Kr~\cite{Raut_2013}. Nuclides shown in grey boxes are stable while those
shown in white boxes are radioactive. The arrows show the s-process path occurring at low (blue) and high (red) neutron densities.}
\label{Branch}
\end{figure}

The $^{86}$Kr(n,2n)$^{85m}$Kr and $^{86}$Kr(n,$\gamma$)$^{87}$Kr reactions can also be useful as test cases in astrophysics.
$^{85}$Kr is an important branching point in the slow-neutron capture process (s-process) that occurs in asymptotic giant branch (AGB) stars, which are thought to produce about half
of all elements heavier than iron~\cite{Raut_2013}.
Branching points in the s-process occur at isotopes with decay half-lives that are long enough to allow the neutron capture reaction to compete with the radioactive decay.
In the case of $^{85}$Kr, with a half-life of 10.75 y, neutron capture is competitive in typical AGB star environments, as shown in Fig. \ref{Branch}.
Of course, the knowledge of the $^{85}$Kr(n,$\gamma$)$^{86}$Kr cross section is extremely important to obtain information on the s-process from the study of
this branching point, but, unfortunately, due to $^{85}$Kr radioactivity, the $^{85}$Kr(n,$\gamma$)$^{86}$Kr reaction has not been measured directly above thermal energies
and only one measurement exists in the thermal region~\cite{bemis_1972}. A possible way to circumvent this problem is to follow the path outlined in Ref.~\cite{Raut_2013},
where the authors were able to estimate the $^{85}$Kr(n,$\gamma$)$^{86}$Kr cross section up to the neutron energy of 10 MeV using the available data of
the $^{86}$Kr($\gamma$,$\gamma^{\prime}$)$^{86}$Kr cross sections~\cite{Schwengner_2013} and performing a series of measurements of the $^{86}$Kr($\gamma$,n)$^{85}$Kr
cross sections in the 10 to 13 MeV $\gamma$-ray energy range. Once a good description of the processes was achieved, they used the same input to calculate the
reverse $^{85}$Kr(n,$\gamma$)$^{86}$Kr process. In this context, the $^{86}$Kr(n,2n)$^{85m}$Kr reaction provides a check for the modeling of the $^{86}$Kr($\gamma$,n)$^{85}$Kr
reaction that has been measured to study the $^{85}$Kr radiative capture cross section.
The $^{86}$Kr(n,$\gamma$)$^{87}$Kr reaction cross section is also important for the calculation of the abundance of $^{87}$Kr isotopes in AGB stars~\cite{Bhike_2015}.
For use in these applications, a new evaluation of the neutron-induced reactions on $^{86}$Kr has been performed using a new resolved resonance region evaluation and the fast
region nuclear reaction code EMPIRE~\cite{empire} which employs several physical models. In particular, the current evaluation also provides
recommended values for $\gamma$-ray production and isomeric cross section production, which are absent in the previous work and are important for the NIF diagnostics and for
astrophysical applications. This is a major improvement from the ENDF/B-VIII.0 evaluation as it did not contain any isomeric nor $\gamma$-production information, and had significant issues in the thermal and resonance regions.

This article is organized as follows: In Section~\ref{Sec:rrr} we describe the evaluation improvements  in the resonance region; in Section~\ref{Sec:fast_data} we detail the
developments done in the fast-neutron range; we present our conclusions in Section~\ref{Sec:conclusions}.

\section{Resolved Resonance Region}\label{Sec:rrr}

We review previous evaluation efforts in the Resolved Resonance Region (RRR) in Section \ref{sec:RRR-history} and then present our updates and improvements in Section~\ref{new_RRR}.

\subsection{Historical Background}
\label{sec:RRR-history}

The RRR provided in the latest release of the ENDF/B library, namely ENDF/B-VIII.0~\cite{Brown:2018}, comes from the
ENDF/B-VII.1~\cite{Chadwick:2011} evaluation, which was fully based on the 2006 edition of the Atlas of Neutron Resonances~\cite{Atlas2006}.
Those resonances are substantially different from the ones found in JENDL-4.0~\cite{JENDL4.0}. The 2006 Atlas is missing the -20 keV artificial level employed by
JENDL-4.0 to represent the low-energy $1/v$ behavior and the correct thermal cross section. The RRR from JENDL-4.0 stops at 650 keV and above this energy, the 2006 Atlas has
many more resonances (74 more, up to 946.5 keV). These high energy resonances are taken from Ref.~\cite{Carlton_1988}. The references above indicate that JENDL-4.0 adopted its
resonances from the 1981 edition of the Atlas~\cite{Atlas1981}, amended with data from Carlton \etal \cite{Carlton_1988} and Raman \etal \cite{Raman_1983}.


The original 1981 Atlas evaluation was used as a basis for the WPEC SG-23 evaluation~\cite{WPEC23}.
Data from Carlton \etal and Raman \etal were then used to revise the evaluation.
JENDL-4.0~\cite{JENDL4.0} adopted a development version of the WPEC SG-23 evaluation before
it could be finalized.  Subsequently,
the Atlas was revised to its 2006 state~\cite{Atlas2006} and that  was used in
the final WPEC SG-23 evaluation. This evaluation eventually became part of ENDF/B-VII.1~\cite{Chadwick:2011}.

\subsection{Current Resonance Evaluation}
\label{new_RRR}

The current evaluation of the total cross section was initially based on fits to separate both data taken by
J. Harvey's team (Refs.~\cite{Carlton_1988,Raman_1983}, EXFOR entries 13149 and 12837) at the Oak Ridge Electron Linear Accelerator (ORELA).
Initially these data were compiled to the Atlas 2006~\cite{Atlas2006} assuming the Multi-level Breit-Wigner (MLBW) approximation.  
There are indications that ORELA resonances were in reality fit with Reich-Moore (RM), rather than MLBW.
When the Atlas 2006 values were reinterpreted as RM parameters,
we noted a dramatic improvement in the agreement between the Atlas RRR parameters and experimental data.
For neutron capture cross sections, the average gamma widths were taken from capture integrals and assigned to resonances from total cross sections.
This explains the lack of dispersion in the capture width distributions. The elastic cross section was defined from the subtraction of capture widths from total widths.

\begin{figure}[hbpt]
\includegraphics[width=1\columnwidth]{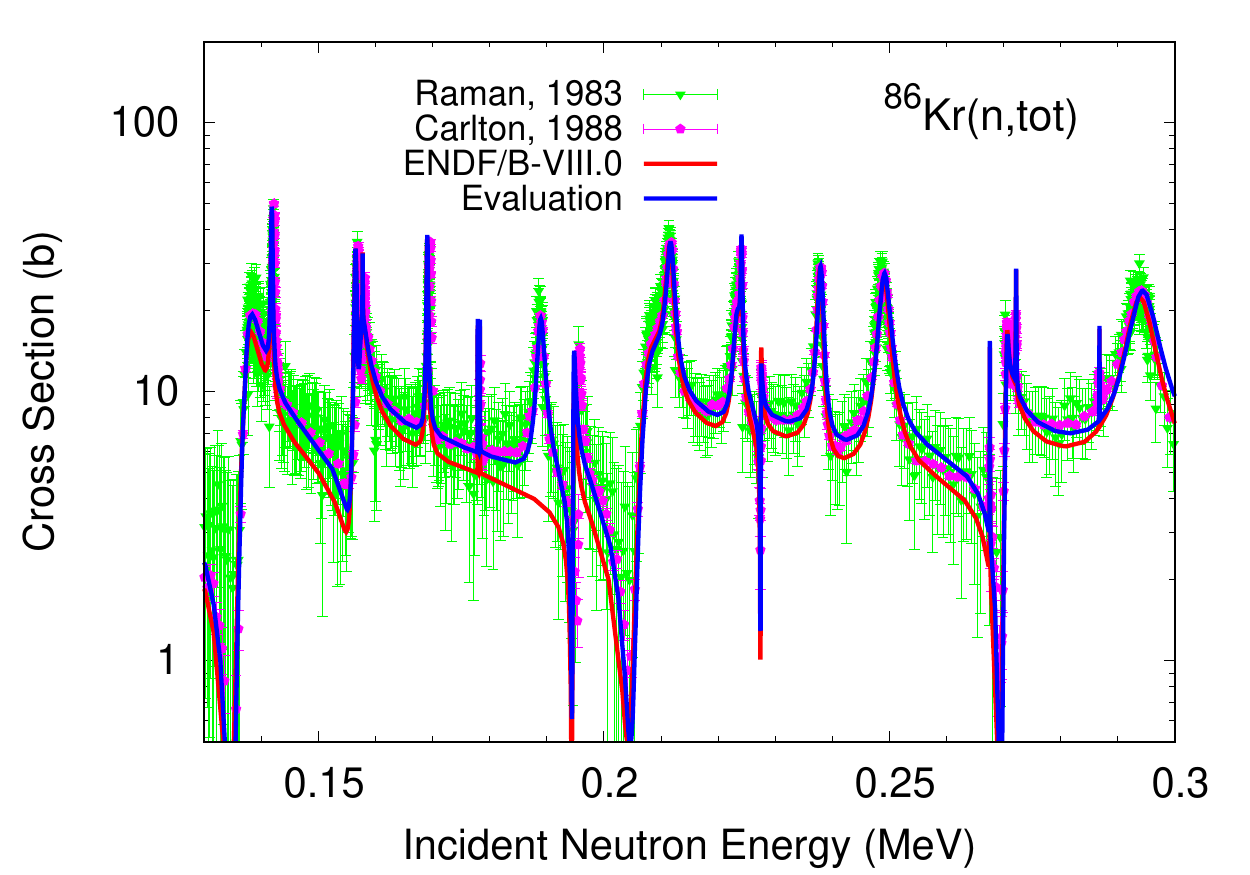}
\caption{Evaluation of the $(\mathrm{n},\mathrm{tot})$ cross section. A missing resonance at 188.89 keV from Carlton \etal~\!\!\cite{Carlton_1988} has been added and the data background has also been corrected.
Experimental data are taken from Refs.~\cite{Carlton_1988,Raman_1983}.}
\label{added_resonance}
\end{figure}

The major changes relative to the ENDF/B-VIII.0~\cite{Brown:2018} resonances, which had been taken from ENDF/B-VII.1~\cite{Chadwick:2011}, can be summarized as follows:
\begin{itemize}
    \item Change of resonance format from MLBW to RM in total cross sections.
    \item Setting the scattering radius $R^{\prime}$ to match the value recommended by the Atlas~\cite{Atlas2018}, that is  7.8 $\pm$ 0.1 fm.
    \item Adding the background artificial resonance at \mbox{-20.0}~keV (adopted from JENDL-4.0~\cite{JENDL4.0}), but with tuned $\Gamma_\gamma$ to set
      thermal capture cross section at the value of 3 mb. At first this may seem low, but it is worth noting that the 3 mb value is based solely on measurements of 4 primary gammas.
    \item Adding a missing resonance at 188.89 keV from Carlton \etal~\!\!\!\cite{Carlton_1988}  using systematics for  $\Gamma_\gamma$ taken from JENDL-4.0~\cite{JENDL4.0} (see
            Fig.~\ref{added_resonance}).
\end{itemize}

\begin{figure}[hbpt]
\includegraphics[width=1\columnwidth]{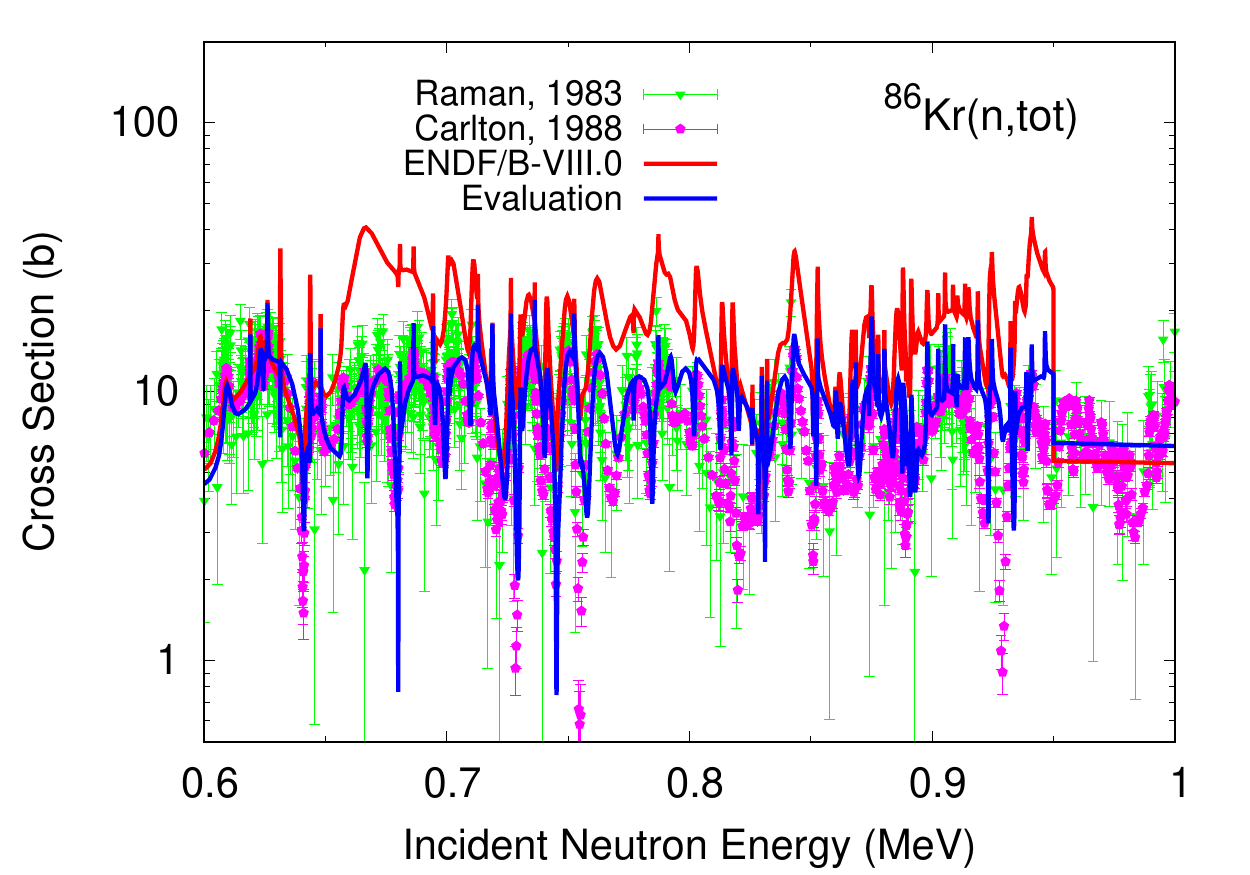}
\caption{Evaluation of the $(\mathrm{n},\mathrm{tot})$ cross section for energies between 600 keV and 1 MeV. This significant improvement arises from the correct interpretation of the widths as Reich-Moore (RM) fits rather than Multi-level Breit-Wigner (MLBW).
Experimental data are taken from Refs.~\cite{Carlton_1988,Raman_1983}.}
\label{total_fixed_background}
\end{figure}

\begin{figure}[hbpt]
\includegraphics[width=1\columnwidth]{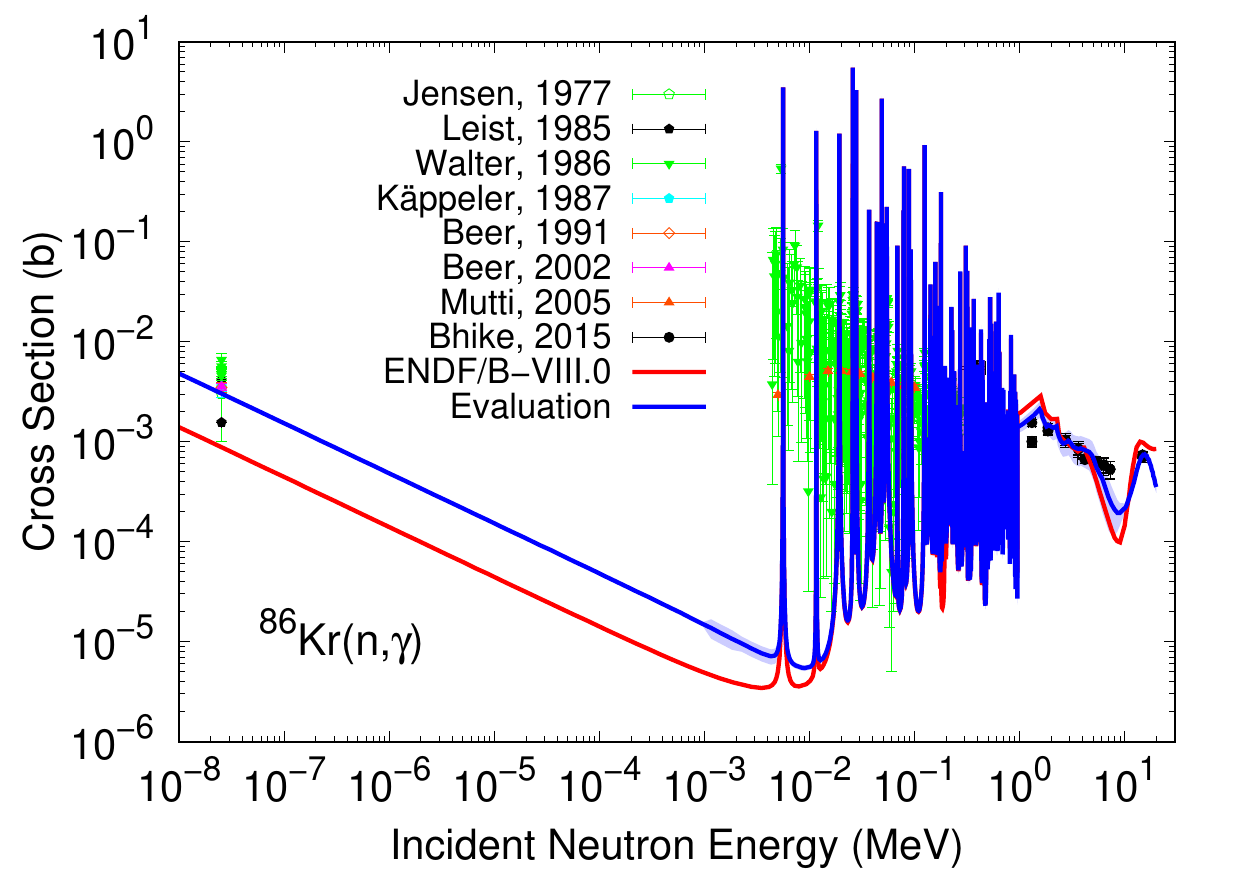}
\caption{Evaluation of the $(\mathrm{n},\gamma)$ cross section compared to ENDF/B-VIII.0~\cite{Brown:2018}.
Experimental data are taken from Refs.~\cite{Bhike_2015,Mutti_2005,Walter_1986,PhysRevC.35.936,Beer_1991,Beer_2002,Leist_1985,PhysRevC.15.1972}.}
\label{capture_xsec_RRR}
\end{figure}

Overall agreement of \nuc{86}{Kr}(n,tot) improved dramatically at high energies on account of the format change, which can be clearly seen in Fig.~\ref{total_fixed_background};
and at low energy mainly due to the new resonance (Fig.~\ref{added_resonance}) and the change to $R^{\prime}$. The \nuc{86}{Kr}(n,$\gamma$) cross section now goes though the thermal point, but is a bit
lower than the average gamma data in experiment (Fig.~\ref{capture_xsec_RRR}).  It is unclear, from the data perspective, whether this could be further improved. There were indications, however, that there may be
missing p-wave resonances around the 188.89 keV resonance, as there seem to be small energy shifts in that region.

In Table~\ref{Tab:integral_metrics} the thermal cross-section values, the resonance integrals (RI), and the Maxwellian averaged cross section (MACS) of $^{86}$Kr are reported for
several evaluated and measured data. We note that the thermal values computed in this evaluation do not have uncertainties. A proper set of covariances in the thermal region and RRR would require a full R-matrix~\cite{RevModPhys.30.257,Breit1959} analysis, which is outside the scope of this work. The uncertainty bands seen in the cross-section plots in the RRR were obtained from extrapolated covariances from EMPIRE~\cite{empire} calculations, which have energy lower-limit of $\sim$1 keV, constrained by fast-region data\footnote{See Section \ref{sec:covariances}.}.

\begin{table}[!htp]
\caption{$^{86}$Kr resonance region integral metrics. Atlas results marked with ``exp'' refer to the Mughabghab’s evaluation based on the experimental data, while Atlas values
derived from calculations are marked with ``calc''.}
\begin{center}
\begin{tabular}{r|ccc}
\toprule \toprule
 $\sigma_{\mathrm{therm}}$ (b) &  ($n$,tot)  & ($n$,el)  & ($n$,$\gamma$)  \\ 
 \midrule
  This work\footnote{Thermal values do not have uncertainties since EMPIRE~\cite{empire} can only generate covariances down to 1 MeV.} & 6.695 & 6.689 & 3.028e-3 \\
  Atlas (2018)~\cite{Atlas2018} & $6.72 \pm 0.37$ & $6.72 \pm 0.37$ & $(3 \pm 2)$e-3  \\
  ENDF/B-VIII.0~\cite{Brown:2018} & 5.442 & 5.441 & 8.790e-4 \\
  JEFF-3.3~\cite{Plompen2020} & 5.824 & 5.818 & 2.968e-3 \\
  JENDL-4.0~\cite{JENDL4.0} & 6.155 & 6.150 & 3.001e-3 \\
 \bottomrule \bottomrule
\end{tabular} \\
\begin{tabular}{r|ccc}
\toprule \toprule
 RI (b) &  ($n$,tot)  & ($n$,el)  & ($n$,$\gamma$)  \\ 
 \midrule
  This work & $114.3 \pm 1.4$ & $109.6 \pm 1.4$ & $(20.4 \pm 1.3)$e-3 \\
  Atlas (2018)~\cite{Atlas2018} &  &  & $(18 \pm 2)$e-3  \\
  ENDF/B-VIII.0~\cite{Brown:2018} & 100.5 & 97.35 & 20.00e-3 \\
  JEFF-3.3~\cite{Plompen2020} & 110.7 & 104.3 & 18.70e-3 \\
  JENDL-4.0~\cite{JENDL4.0} & 105.1 & 101.2 & 23.42e-3 \\
 \bottomrule \bottomrule
\end{tabular} \\
\begin{tabular}{r|ccc}
\toprule \toprule
 MACS(30 keV) (mb) &  ($n$,tot)  & ($n$,el)  & ($n$,$\gamma$)  \\ 
 \midrule
  This work & $7880 \pm 310$ & $7870 \pm 310$ & $5.08 \pm 0.28$ \\
  Atlas (exp) (2018)~\cite{Atlas2018} &  &  & $4.76 \pm 0.28$  \\
  Atlas (calc) (2018)~\cite{Atlas2018} &  &  & $5.3 \pm 0.5$  \\
  ENDF/B-VIII.0~\cite{Brown:2018} & 6639 & 6634 & 5.063 \\
  JEFF-3.3~\cite{Plompen2020} & 6977 & 6973 & 4.084 \\
  JENDL-4.0~\cite{JENDL4.0} & 7335 & 7330 & 5.112 \\
  Kadonis-1.0\footnote{This value is taken from Ref.~\cite{Mutti_2005} where the authors used Xenon as a reference.}~\cite{kadonis1.0} & & & $4.76 \pm 0.28$ \\
  \midrule
  K\"appeler (2021)~\cite{kaeppeler_2021} & & & $4.8 \pm 0.7$ \\  
  Mutti (2005)~\cite{Mutti_2005} & & & $4.76 \pm 0.28$ \\
  Beer (1991)~\cite{Beer_1991} & & & $3.34 \pm 0.24$ \\
  Walter (1986)~\cite{Walter_1986} & & & $5.6 \pm 0.7$ \\
  Walter (1986)~\cite{Walter_1986_2} & & & $3.8 \pm 0.7$ \\
  Raman (1983)~\cite{Raman_1983} & & & $4.8 \pm 1.2$ \\
 \bottomrule \bottomrule
\end{tabular}
\end{center}
\label{Tab:integral_metrics}
\end{table}

\section{Evaluation in the Fast Range}
\label{Sec:fast_data}

In Section~\ref{sec:fast-data} we discuss the experimental data in the fast range available to provide constraints on model calculations.  These calculations are discussed in Section~\ref{Sec:eval}. The evaluated results are then presented in Section~\ref{subSec:x-sec}.

\subsection{Availability of Experimental Data}
\label{sec:fast-data}

In the fast region we have data for only four processes: total, inelastic $\gamma$-transitions, $(\mathrm{n},2\mathrm{n})$, and radiative capture.
The $(\mathrm{n},2\mathrm{n})$ channel
opens up at an energy of $\sim 10$ MeV and can be considered at a later stage during the evaluation.
Also, the radiative capture cross section is much smaller compared to the cross section of the other processes, so it can be considered in a later step of the evaluation.

For the total cross section we have two data sets:
one from Raman \etal~\!\!\!\cite{Raman_1983}, which covers the energy range that goes from $4$ keV to $5$ MeV, and another one measured by
Carlton \etal~\!\!\cite{Carlton_1988} covering a wider energy range that goes from $1$ MeV to $25$ MeV. For this latter set, some considerations have to be made.
Due to poor statistics, it provides different levels of reliability depending on the incident energy of the incoming neutron. In particular, in the energy range between $1$ and
$8$ MeV the average cross section has small error bars and thus can be trusted with a higher degree of confidence. In contrast, for higher energies the uncertainties become larger, but it is still possible to
identify an intermediate range, from $8$ to $12$ MeV, where the average cross section has still relatively small error bars and the data can still be trusted. Above $12$ MeV the uncertainties and fluctuations
in the average cross section are too large and it is not possible to rely on the data anymore.

In addition to the $(\mathrm{n}, \mathrm{tot} )$ data we also have precise data for $12$ inelastic transitions~\cite{Fotiades_2013} that we can use to constrain our model.
With a simple look at the data one realizes that the cross sections for the different transitions are in general of the same order of magnitude except for two that are larger.
In particular, the transition from the first excited state ($2^+$) to the ground state  ($0^+$) is dominant with one order of magnitude of difference with respect to the second most important
transition and almost two orders of magnitude with respect to all the other ones.
Based on these considerations we adopted the following strategy: we performed two different fits of the $(\mathrm{n}, \mathrm{tot} )$ data in the first two energy regions, {\it i.e.} from
$0.5$ to $8$ MeV and from $8$ to $12$ MeV, and we fitted our optical potential to simultaneously reproduce these fits and the data of the $2^+ \rightarrow 0^+$ inelastic transition
for energies up to $11$ MeV. For energies above $12$ MeV we adopted the pure Koning-Delaroche~\cite{Koning:2003} optical potential and we performed a smooth transition
between the fitted potential and the pure Koning-Delaroche potential in the energy region between $11$ and $12$ MeV.
Finally, for energies above $12$ MeV the description of the $(\mathrm{n} , 2\mathrm{n})$ data should be dictated by the pre-equilibrium, so for this energy range we tuned the
pre-equilibrium model to simultaneously reproduce the $(\mathrm{n} , 2\mathrm{n})$ data and the high-energy inelastic data.


\subsection{Reaction Model Parametrizations}
\label{Sec:eval}
\begin{figure}[hbpt]
\begin{center}
\includegraphics[scale=0.37,clip,trim= 2mm 7mm 7mm 0mm]{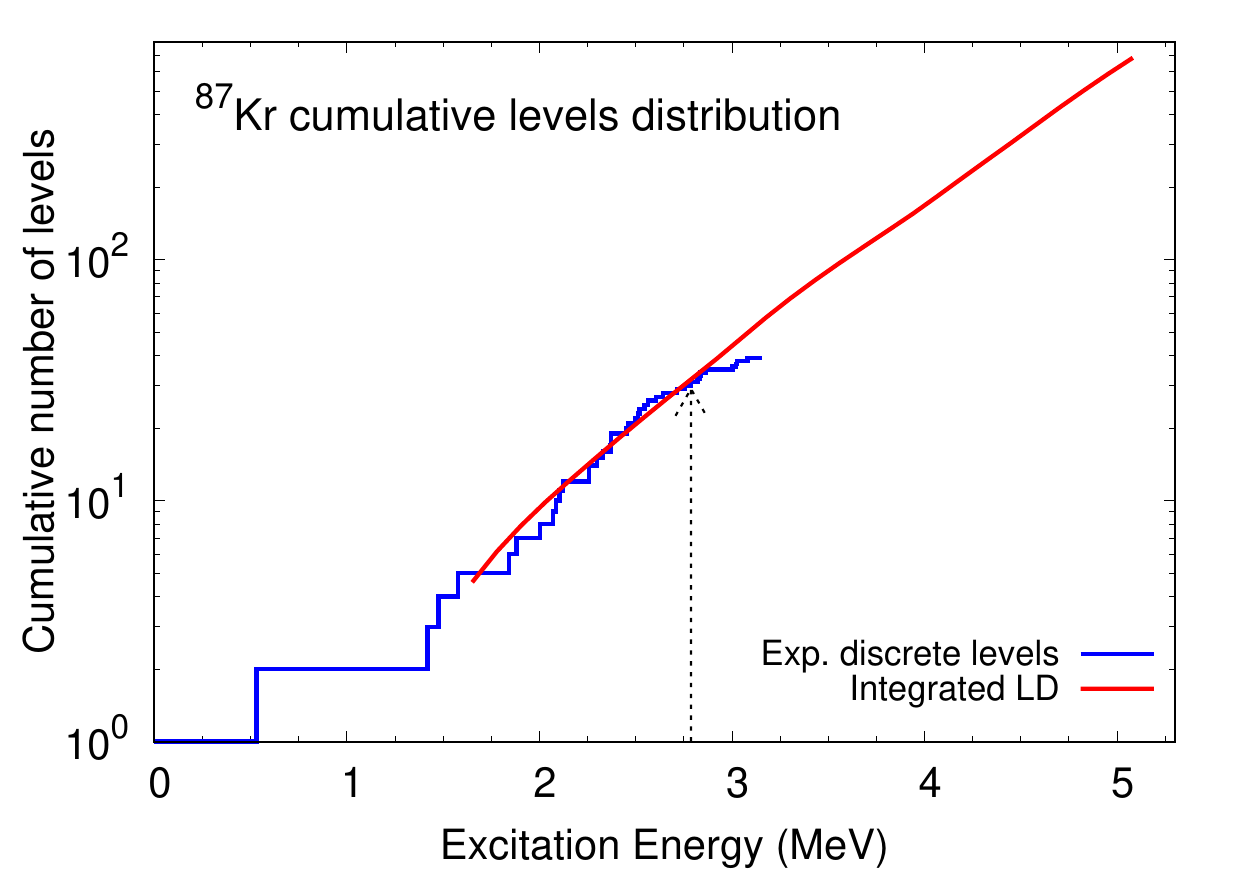}
\includegraphics[scale=0.37,clip,trim= 7mm 7mm 7mm 0mm]{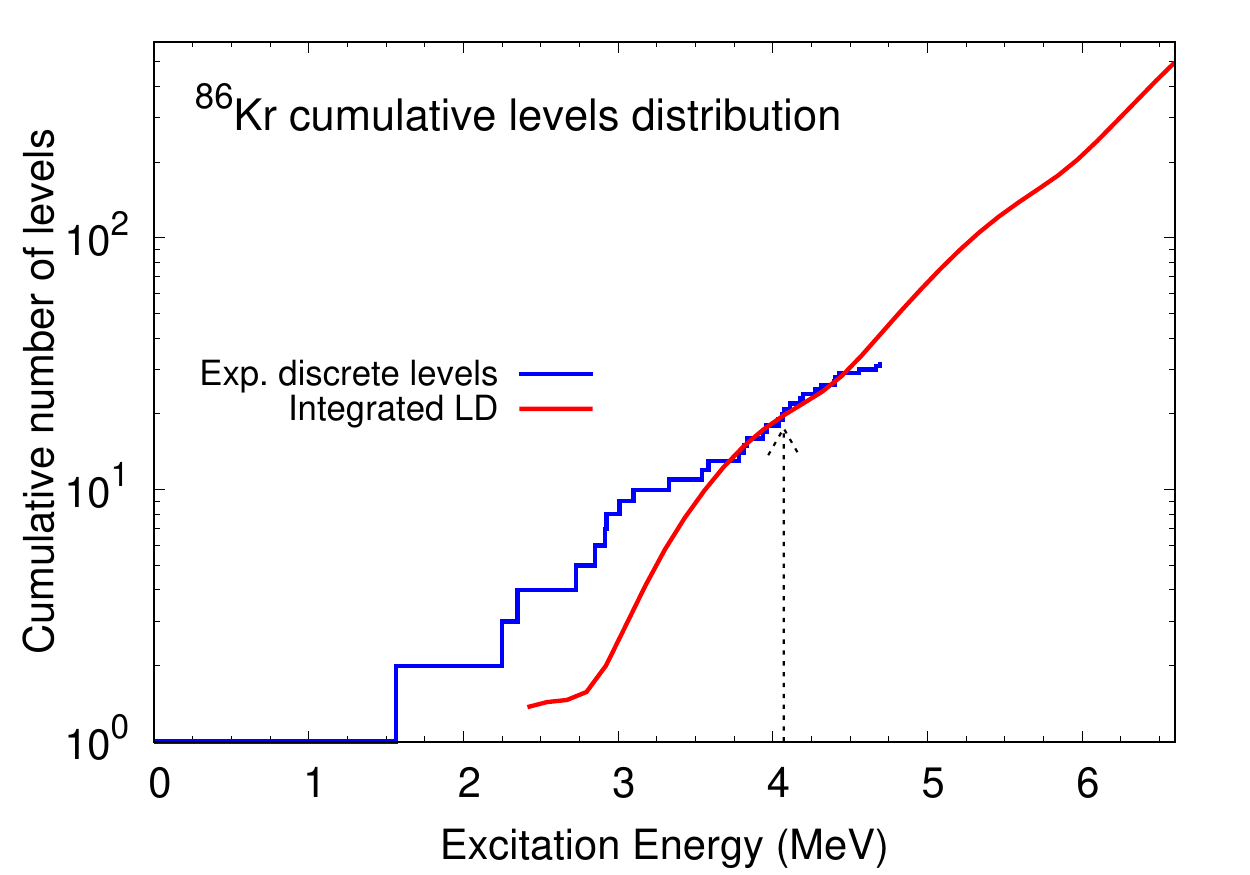}
\includegraphics[scale=0.37,clip,trim= 2mm 0mm 7mm 0mm]{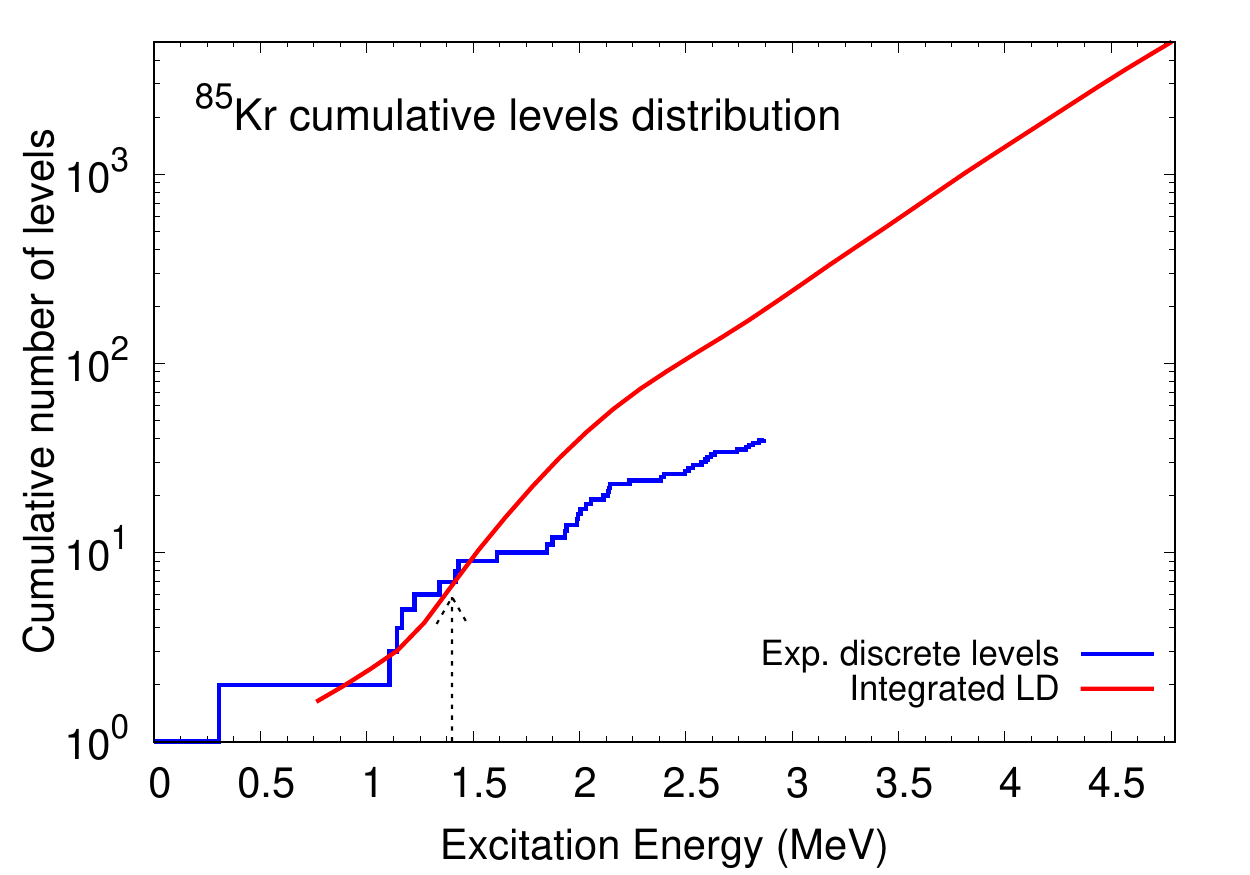}
\includegraphics[scale=0.37,clip,trim= 7mm 0mm 7mm 0mm]{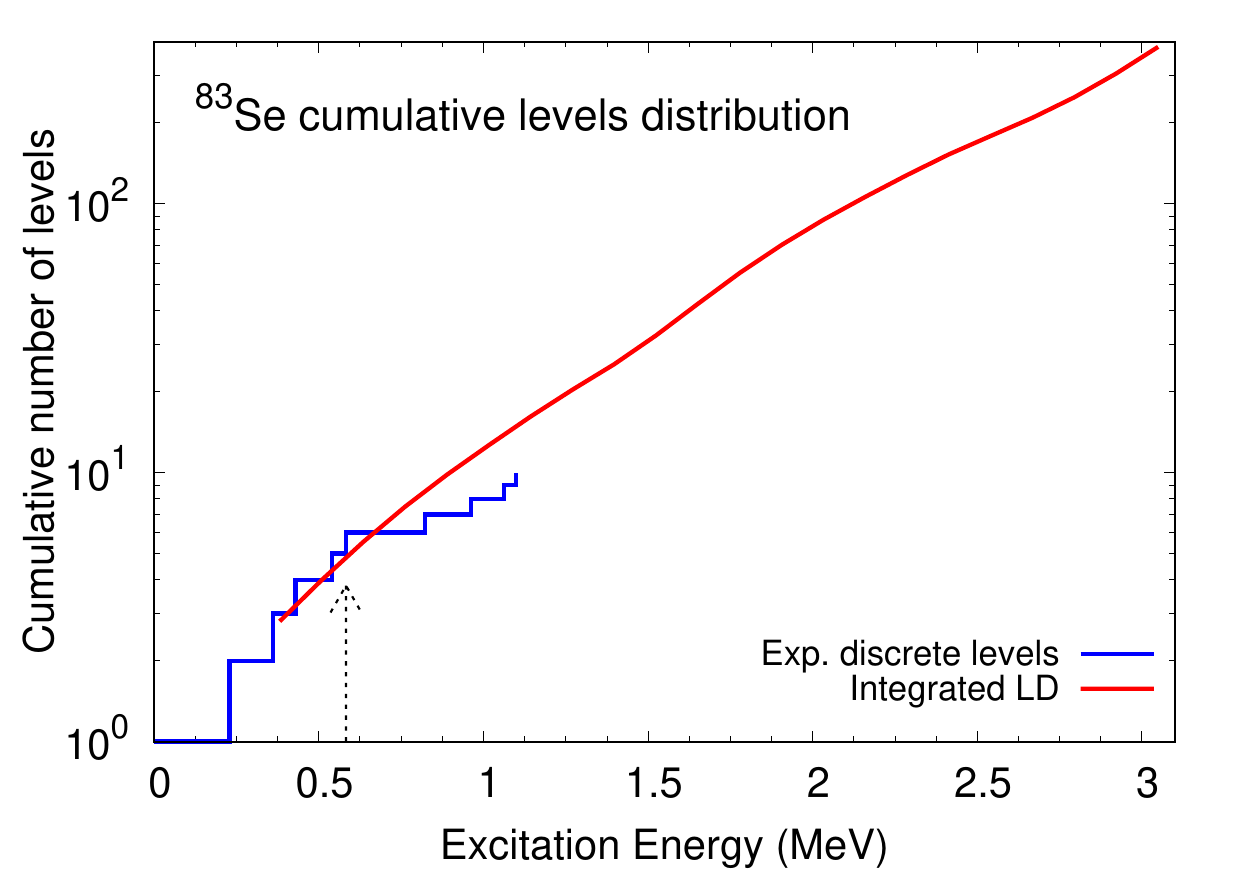}
\caption{Cumulative level distribution obtained from the levels listed in RIPL-3~\cite{RIPL3} (blue lines) compared with the level densities employed (red curves) in our model calculations
for the residual nuclei from the main reactions studied in this evaluation work. Dashed arrows show the discrete-level cutoff adopted indicating transition to level continuum.}
\label{fig:CLD}
\end{center}
\end{figure}

We employed the EMPIRE nuclear reaction code \cite{empire} for model calculations and parameter tuning aiming to evaluate the reaction channels in the fast region (i.\!\!\!\! e., neutron incident energies above the first inelastic excitation energy).
The level-density (LD) model adopted was the combinatorial Hartree-Fock-Bogoliubov (HFB) from \mbox{RIPL-3} \cite{RIPL3}, as its microscopic nature should in principle provide more realistic spin and parity distributions for  a nucleus with such scarce availability of data such as \nuc{86}{Kr}.  The lack of differential spectra data prevented us from using the approach of Ref.~\cite{Nobre:2020} to fine tune and better constrain the LD, as it has been done recently in Ref.~\cite{Nobre_2021}.
We adjusted the LD to optimize the agreement with observed discrete levels. Fig.~\ref{fig:CLD} shows the experimental discrete levels compared to the cumulative level distributions obtained from the adopted LD  for the residual nuclei of the relevant reactions for the current evaluation. The continuum energy cut-off $E_{\mathrm{cut}}$, above which levels are considered to be in the continuum and calculations switch from discrete levels to LD, were 2.787, 4.070, 1.400, and 0.582~MeV for \nuc{87}{Kr}, \nuc{86}{Kr}, \nuc{85}{Kr}, and \nuc{83}{Se}, respectively, and are marked as dashed arrows in the plots. We can see that, generally, there is a very good agreement of the distribution and its shape with the the discrete levels at and around $E_{\mathrm{cut}}$.

Width fluctuation corrections were applied up to 9.0~MeV and pre-equilibrium was determined using the exciton model~\cite{exciton} as implemented in the code PCROSS~\cite{PCROSS}, with mean free-path, single-particle LD parameter, and coefficient defining the equilibrium exciton number adjusted to reproduce the experimental cross-section data.
The complete $\gamma$-ray cascades were calculated and  we assumed  the RIPL-3 MLO1 formulation by Plujko for E1 transitions initiated from the continuum for the $\gamma$-strength functions of all residual nuclei produced~\cite{RIPL3,MLO1}.
Electromagnetic transitions of multipolarity M1, E1 and E2 were considered.  The decay from the compound nucleus was added incoherently to the direct cross sections.
 Blatt-Biedenharn coefficients were employed to calculate compound-nucleus anisotropy~\cite{BlattB}.

In Table~\ref{Tab:omp-ripl}, we identify the  optical model potentials (OMPs) employed for the different reaction channels. For the direct reaction component, due to the lack of recent local OMP specific to \nuc{86}{Kr}, we adopted the spherical OMP of Koning-Delaroche.
One of the main experimental constraints for this evaluation are cross-section data from gamma transitions between inelastic levels, which is highly dependent on level-coupling mechanisms.  Therefore, we opted to describe explicitly some of these couplings.
For that, we employed coupled-channel calculations within a rigid rotational model, coupling to the first 2$^{+}$ and 4$^{+}$ states using a quadrupole deformation of $\beta_2=0.145$ \cite{RIPL3,Raman:2001} and the Koning-Delaroche as the bare potential.
This allows for minor tuning of total and absorption cross sections following the data available. This corresponds to an intermediate approach between spherical nuclei and adiabatic approximations for more strongly-deformed nuclei \cite{Nobre:2015}, which is consistent with the intermediate deformation seen in \nuc{86}{Kr}.
Levels computed in the distorted-wave Born approximation (DWBA) were considered in calculations up to the transition to continuum at 4.07~MeV in order to improve agreement with inelastic gamma data and to generate neutron production spectra without gaps in the transition to continuum/preequilibrium contributions.

\begin{table}[!htp]
\caption{Optical model potentials used in the EMPIRE~\cite{empire} calculations. The ``CC'' label indicates a coupled-channels optical model potential, in contrast with the spherical ones.}
\begin{tabular}{lccr}
\toprule \toprule
 Ejectile        & Type   & RIPL \#       & Reference \\ 
 \midrule
 \multirow{2}{*}{$n$ (direct)}  & CC & \multirow{2}{*}{ 2405       }                 & \multirow{2}{*}{Koning+~\cite{Koning:2003}}    \\
                                              & (rotational model)                &                                &                 \\
 $n$ (compound)  & Spher.                    &  2405       &  Koning+~\cite{Koning:2003} \\
 $p$                      & Spher.                    &  5405       &  Koning+~\cite{Koning:2003} \\
 $\alpha$              & Spher.                    &  9600       &  Avrigeanu+~\cite{Avrigeanu:1994}  \\
 $d$                      & Spher.                    &  6200       &  Haixia+~\cite{Haixia:2006}  \\
 $t$                       & Spher.                    &  7100       &  Becchetti+~\cite{Becchetti-Greenlees}  \\
 \nuc{3}{He}          & Spher.                    &  8100       &  Becchetti+~\cite{Becchetti-Greenlees} \\  
 \bottomrule \bottomrule
\end{tabular}
\label{Tab:omp-ripl}
\end{table}

\subsection{Results}
   \label{subSec:x-sec}

We present here the main results for the current evaluation. Whenever data is not shown, it can be assumed there are no experimental data available in the literature for that particular reaction and/or energy range. Experimental data, when present, were obtained through the experimental nuclear reaction data (EXFOR)~\cite{OTUKA2014272,ZERKIN201831} web utility.

 \subsubsection{Covariances}
 \label{sec:covariances}

The lack of sufficient experimental data and detailed documentation of uncertainty sources prevented us from developing a more quantitative, higher fidelity set of covariances matrices. For this reason we provide lower-fidelity covariances in the fast region obtained through model parameter variations using the KALMAN code~\cite{Kawano:1997yi}. This was done by tuning the weight of experimental constraints until the uncertainties for each reaction were qualitatively compatible with the expected cross-section uncertainties that can be inferred from the actual data. We extended these covariances down to 1~keV.

\subsubsection{Total Cross Section}
\begin{figure}[hbpt]
\includegraphics[width=1\columnwidth]{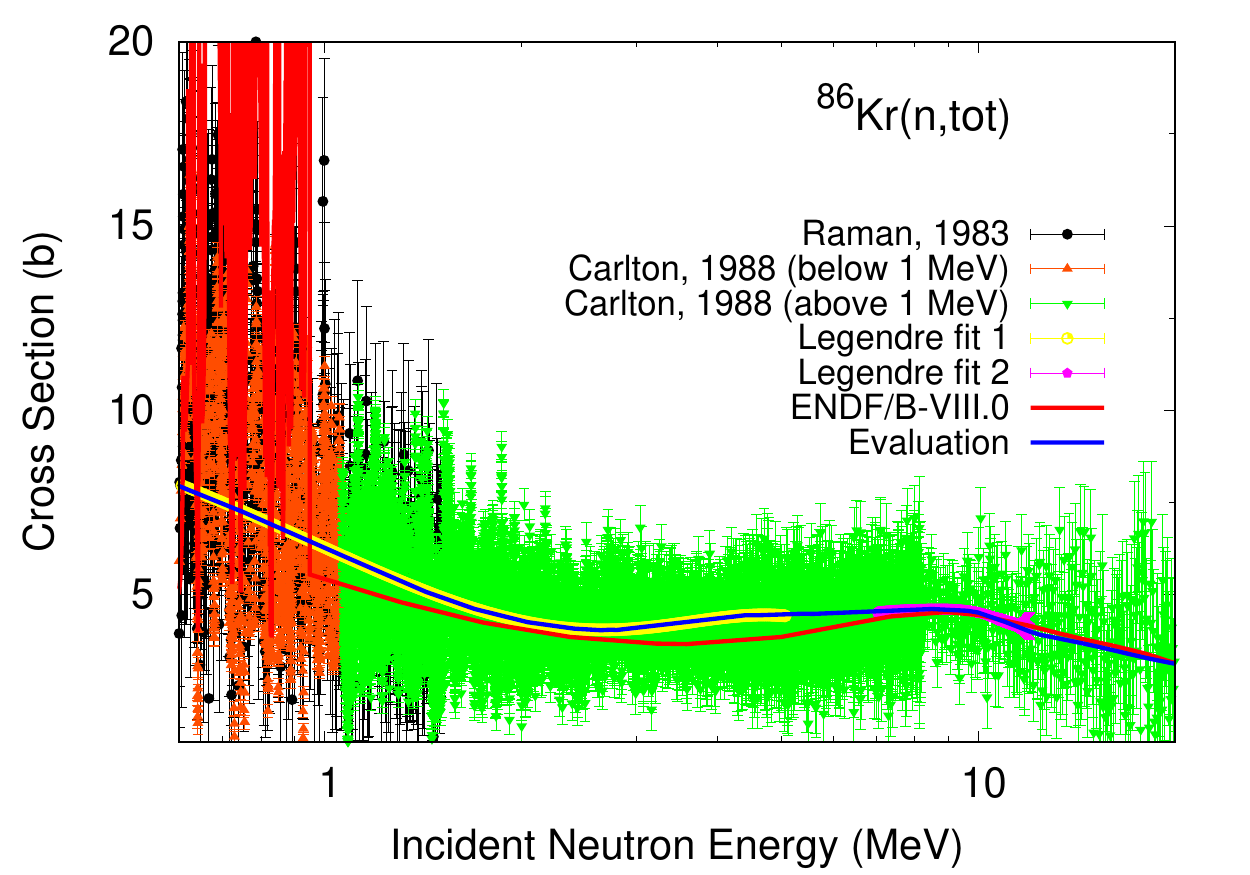}
\caption{Evaluation of the $(\mathrm{n},\mathrm{tot})$ cross section computed with EMPIRE~\cite{empire} and compared with the two fits of the experimental data.
Due to different statistics we performed two different fits of the data for two energy regions 0.5-8 MeV and  8-12 MeV.
Experimental data are taken from Refs.~\cite{Carlton_1988,Raman_1983}.
Ref.~\cite{Carlton_1988} reports different averages for data below and above 1 MeV, and in this work we maintained this distinction.}
\label{total_xsec}
\end{figure}
\begin{figure}[t]
\includegraphics[width=1\columnwidth]{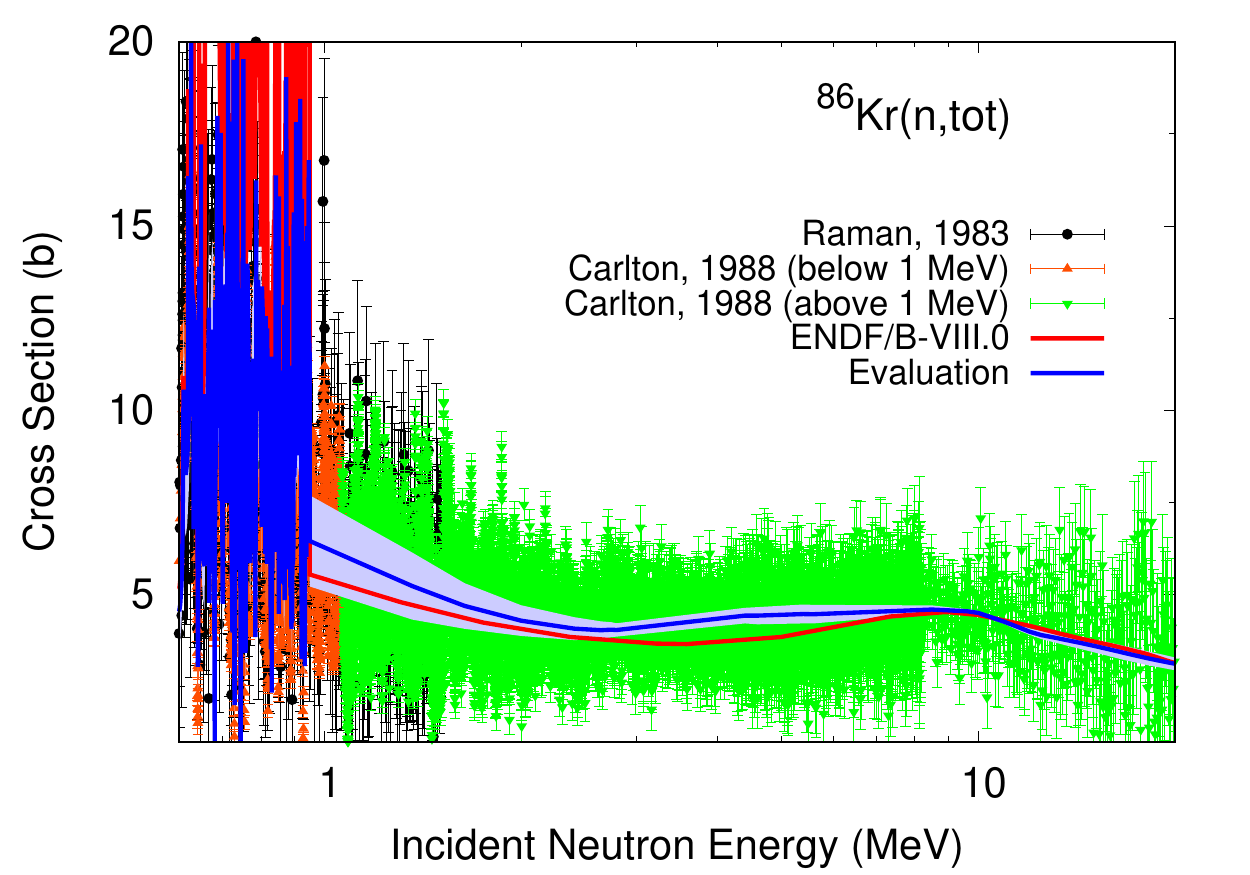}
\caption{Evaluation of the $(\mathrm{n},\mathrm{tot})$ cross section computed with EMPIRE~\cite{empire} and compared with the experimental data~\cite{Carlton_1988,Raman_1983}.
The light-blue band represents the uncertainty band obtained from the covariance matrix.
Ref.~\cite{Carlton_1988} reports different averages for data below and above 1 MeV, and in this work we maintained this distinction.}
\label{total_xsec_2}
\end{figure}
Two measurements of the total cross section of $^{86}$Kr were performed by transmission at ORELA. In 1983 Raman \etal measured
the total cross section in the energy region between $4$ keV and $5$ MeV~\cite{Raman_1983}, and in 1988, Carlton \etal measured the cross section in
the energy region between $1$ MeV and $25$ MeV~\cite{Carlton_1988}.
The experimental data were energy averaged because of the overwhelming number of data points,
to provide a more appropriate comparison with the calculated cross sections\footnote{The computed optical model and Hauser-Feshbach cross sections are energy averages of the fluctuating compound nuclear cross sections.}.
In Fig.~\ref{total_xsec} we display the EMPIRE~\cite{empire} calculation of the total cross section compared with the two fits (yellow and purple lines) of the experimental data that are very nicely
reproduced. The same calculation is also presented in Fig.~\ref{total_xsec_2} along with the resonances and the uncertainty band of the total cross section obtained from the
covariance matrix.

\subsubsection{Inelastic Scattering Cross Sections}
The inelastic scattering reaction provides valuable insight into the nuclear structure of the target nucleus. The compound nucleus decays by emission of a neutron, leaving an
excited $^{86}$Kr nucleus that de-excites by gamma emission. The inelastic scattering gammas are the energy given off as the nucleus transitions from one level to another,
and therefore can be used to study the nuclear structure. Partial gamma cross sections for thirteen such transitions were measured by
Fotiades \etal~\!\!\cite{Fotiades_2013} for incident neutron energies up to $20$ MeV.

The $^{86}$Kr$(\mathrm{n} , \mathrm{n}^{\prime} \gamma) ^{86}$Kr experiment was conducted at the Los Alamos Neutron Science Center (LANSCE) facility using the
Germanium Array for Neutron Induced Excitations (GEANIE) detector system. GEANIE consists of $10$
Compton-suppressed planar Germanium detectors and $10$ Compton-suppressed coaxial Germanium detectors, but the results of the experiment only come from 4 of
the latter detectors chosen for optimal time and energy resolution. The neutron flux was measured using a fission chamber containing $^{238}$U and $^{235}$U foils.
There were ten new transitions discovered that had not previously been observed for this reaction and overall, they measured partial cross sections for $22$ transitions at neutron
energies between $1$ MeV and $20$ MeV. The neutron beam was pulsed and the experimenters used the time-of-flight method to determine the incident energy of the neutrons on target.
The experimenters used natural iron foils on either side of krypton cell, perpendicular to the beam, so the known cross section at $14.5$ MeV incident energy of the $846$ keV
$2^+ \rightarrow 0^+$ could be used for normalization of the cross sections in the krypton experiment. The experimenters also accounted for the efficiency of the detectors,
target thickness, and ``dead-time'' of the neutron beam. More experimental details can be found in Ref.~\cite{Fotiades_2013}.

In an effort to properly reproduce the partial gamma cross section we had to provide branching ratios and $\gamma$-ray energies for several levels that contained incomplete or uncertain structure data. Without these corrections, there would be ambiguity in the decay and EMPIRE~\cite{empire} would assume that the levels in question decay straight to the ground state.
Thus, it was necessary to determine the possible $\gamma$-rays, multipolarities, 
spins, parities, and branching ratios for each energy level with
unspecified decay data in $^{86}$Kr using selection rules and systematics. In Table~\ref{Tab:adopted_Jpi} we display all the discrete levels with uncertain $J^{\pi}$ assignment and the
adopted values that we used to perform our calculations. Similarly, in Table~\ref{Tab:new_gammas} we display all the $\gamma$-rays that we introduced to avoid direct transitions to the
ground state. It is worth notice that in this work we adopted the convention of representing the ground state with the level index 1,
the first excited state with the index 2, and so forth. 

\begin{table}[!htp]
\caption{Adopted $J^{\pi}$ values for discrete levels of $^{86}$Kr. 
In this work we adopted the convention of representing the ground state with the index 1, the first excited state with the index 2, and so forth. All the energies are in MeV.}
\begin{tabular}{cccc}
\toprule \toprule
 Level &  E  & ENSDF $J^{\pi}$  &  Adopted $J^{\pi}$  \\ 
 \midrule
  6 & $2.85072$  & $(2,3)^+$ &  $2^+$ \\
  7 & $2.91683$  & $(3^- )$ &  $3^+$ \\
  9 & $3.00943$  & $(1,2)^+$ &  $2^+$ \\
 11 & $3.32810$  & $(3^+ , 4^+ )$ &  $4^+$ \\
 13 & $3.58340$  & $(0^+ \, \mathrm{to} \, 4^+)$ &  $4^+$ \\
 15 & $3.81632$  & $(5^+ )$ &  $5^+$ \\
 17& $3.93530$  & $(5)$ &  $5^+$ \\
 18 & $3.95900$  & $(3^- , 4^+)$ &  $4^+$ \\
 19 & $4.03860$  & $(2,3)^- $ &  $3^-$ \\
 20 & $4.06412$  & $(6^+ )$ &  $6^+$ \\
 \bottomrule \bottomrule
\end{tabular}
\label{Tab:adopted_Jpi}
\end{table}

\begin{table}[!htp]
\caption{New $\gamma$-rays introduced. In this work we adopted the convention of representing the ground state with the index 1,
the first excited state with the index 2, and so forth. All the energies are in MeV. The subscript $i$ stands for ``initial'' while $f$ stands for ``final''.
The last column shows the branching ratios that we assumed for these transitions.}
\begin{tabular}{cccccccc}
\toprule \toprule
 $\mathrm{Level}_i$ & $\mathrm{E}_i$ & $J_i^{\pi}$ & $\mathrm{E}_{\gamma}$ & $\mathrm{Level}_f$ & $\mathrm{E}_f$ & $J_f^{\pi}$ & BR  \\
 \midrule
 $11$ & $3.3281$  & $4^+$ & $1.078$ & $3$ & $2.25001$ & $4^+$ & $0.4$ \\
 $16$ & $3.8320$  & $0^+$ & $0.823$ & $9$ & $3.00943$ & $2^+$ & $0.4828$ \\
 $16$ & $3.8320$  & $0^+$ & $1.483$ & $4$ & $2.34947$ & $2^+$ & $0.05823$ \\
 $16$ & $3.8320$  & $0^+$ & $2.267$ & $2$ & $1.56461$ & $2^+$ & $0.4589$ \\
 $18$ & $3.9590$  & $4^+$ & $0.631$ & $11$ & $3.32810$ & $4^+$ & $0.04030$ \\
 $18$ & $3.9590$  & $4^+$ & $0.860$ & $10$ & $3.09885$ & $3^-$ & $0.01197$ \\
 $18$ & $3.9590$  & $4^+$ & $1.042$ & $7$ & $2.91683$ & $3^+$ & $0.02061$ \\
 $18$ & $3.9590$  & $4^+$ & $1.610$ & $4$ & $2.34947$ & $2^+$ & $0.01579$ \\
 $18$ & $3.9590$  & $4^+$ & $1.709$ & $3$ & $2.25001$ & $4^+$ & $0.8018$ \\
 $18$ & $3.9590$  & $4^+$ & $2.394$ & $2$ & $1.56461$ & $2^+$ & $0.1095$ \\
 \bottomrule \bottomrule
\end{tabular}
\label{Tab:new_gammas}
\end{table}

For some levels used in our calculations, a specific $J^{\pi}$ assignment was not given in the Evaluated Nuclear Structure Data Files~\cite{Negret:2015,ensdf_2021}
(ENSDF) and a range of possible assignments was given. For some of these levels it was possible to provide a unique $J^{\pi}$ assignment using the values provided in Table IV
of Ref.~\cite{urban_2016}, which reports the results of a subsequent measurement performed after the ENSDF evaluation where some of these
$J^{\pi}$ assignments were better constrained through angular correlation measurements. We also used the information contained in Ref.~\cite{urban_2016} to improve our
transition scheme with the addition of new transitions not included in ENSDF.
Despite that, we still had to provide unique $J^{\pi}$ quantum numbers for some levels. This is the case of levels 9, 11, and 13.
The ENSDF $J^{\pi}$ assignments are determined based on radioactive decay, inelastic scattering measurements, stripping reactions and known gamma
transitions.  For these three levels there was no conclusive choice given, and so in the original EMPIRE~\cite{empire} run, we selected the higher-spin option.
For level 7, a single $J^{\pi}$ assignment was given in ENSDF, but experiments had suggested other possibilities as well.
The $J^{\pi}$ assignments for these four levels were changed based on patterns in the energy dependence of the experimental data, and for each level a new assignment was found
which produced calculated cross sections with energy dependencies that fit the experimental data.
For level 11, the change in spin opened up a new M1 gamma transition of comparable strength with the known transition, and so the branching
ratio for this decay route was calculated using the Weisskopf single-particle estimates~\cite{OakRidgeMemo} and systematics
for the average B(E1), B(E2), and B(M1) for isotones with N=50~\cite{NUDAT}. The mean values of B(E1), B(E2) and B(M1) were found to
be 6.823 $\pm$1.450, 9.398$\times 10^{-7} \pm$ 2.534$\times 10^{-4}$, and 0.2046 $\pm$ 0.1442 respectively.
For all the other levels we used the assignments provided by RIPL without any change.

\begin{figure}[hbpt]
\includegraphics[width=1\columnwidth]{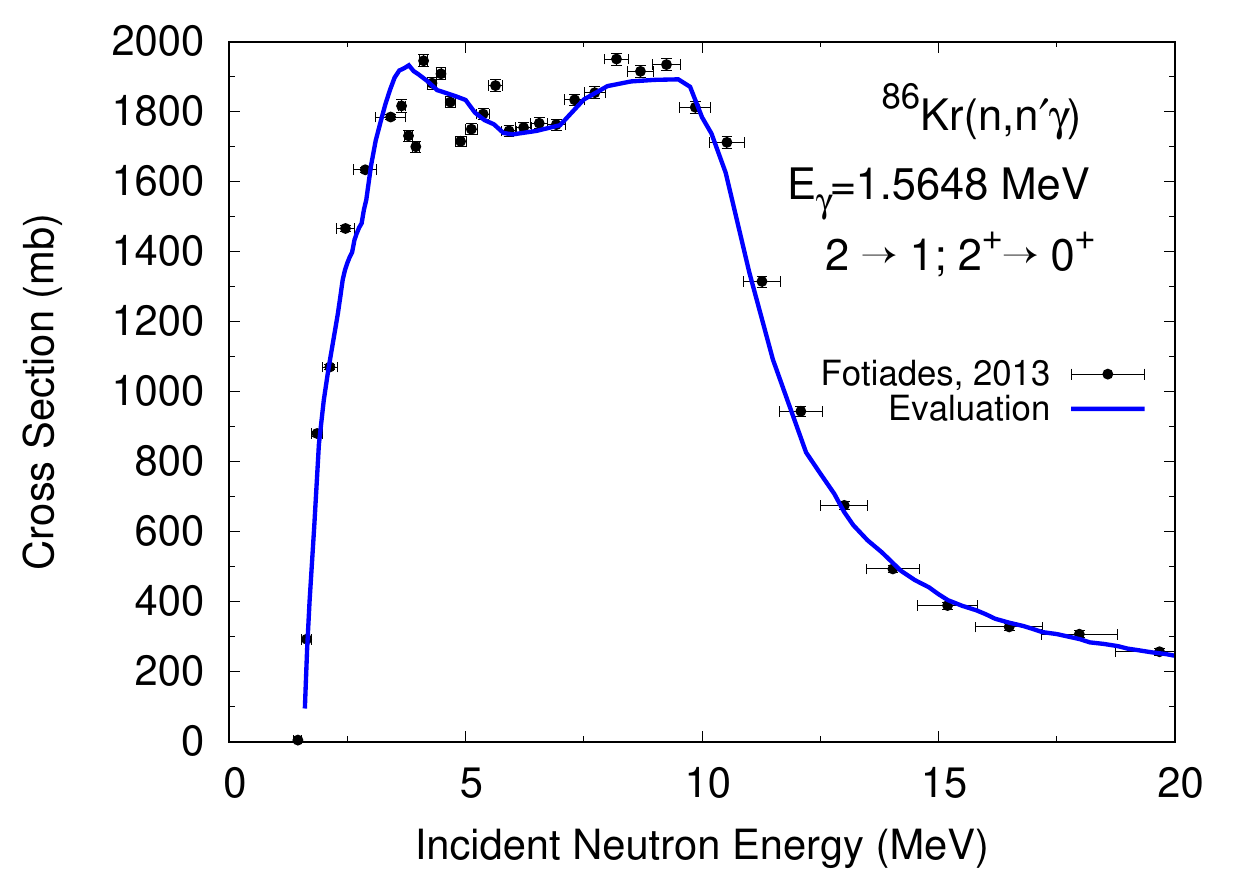}
\caption{Evaluation of the inelastic gamma cross section for the discrete transition from the $2^+$ state to the $0^+$ state.
The indices for initial and final levels of the transition, with their corresponding spin and parity, as well as the $\gamma$ energy, $E_\gamma$, are listed in the plot.
In this work we adopted the convention of representing the ground state with the index 1, the first excited state with the index 2, and so forth.
Experimental data are taken from Ref.~\cite{Fotiades_2013}.  We do not show ENDF/B-VIII.0 here as inelastic gamma information was not present in that evaluation.}
\label{first_inel_transition_xsec}
\end{figure}

\begin{figure*}[hbpt]
\subfloat{\includegraphics[scale=0.51, clip, trim = 0mm 7mm 4mm 2mm]{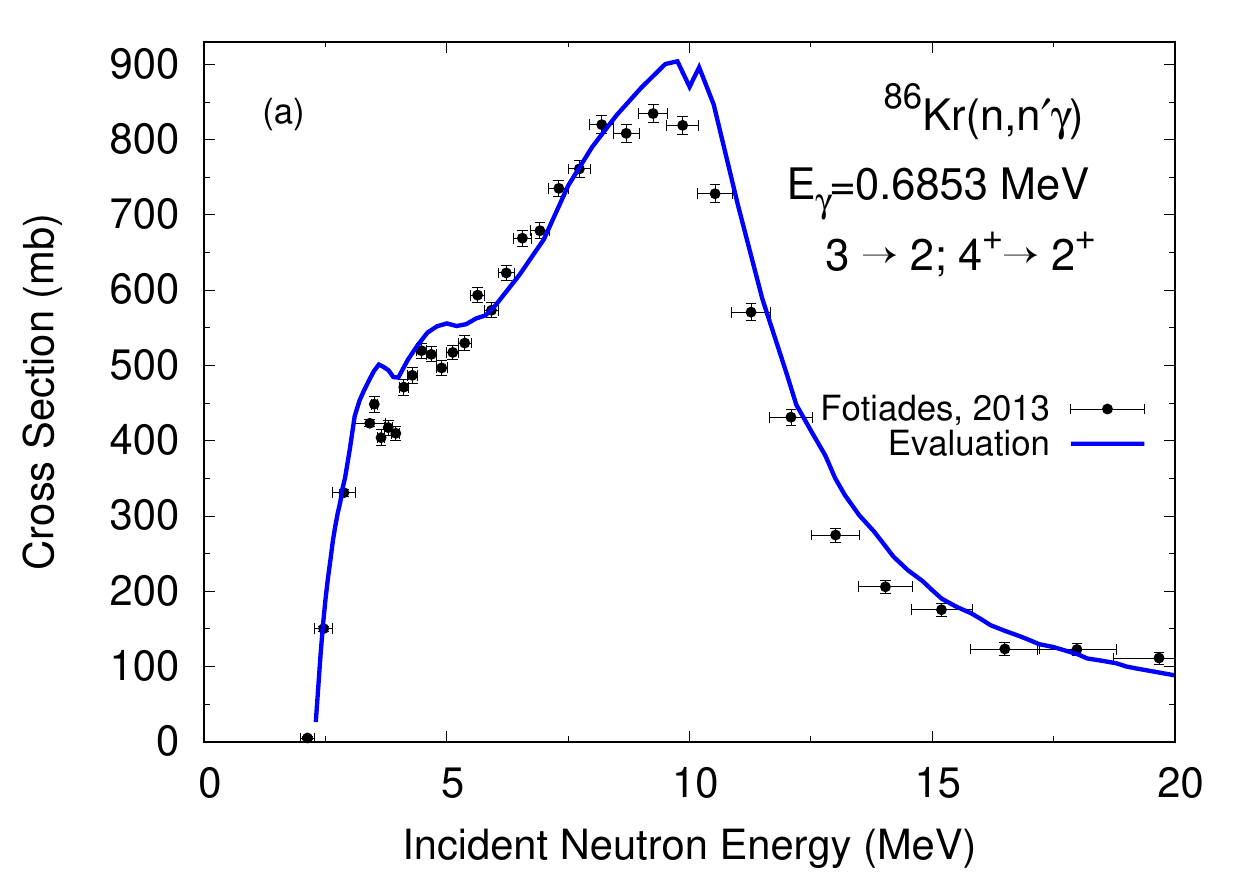}}
\subfloat{\includegraphics[scale=0.51, clip, trim = 9mm 7mm 4mm 2mm]{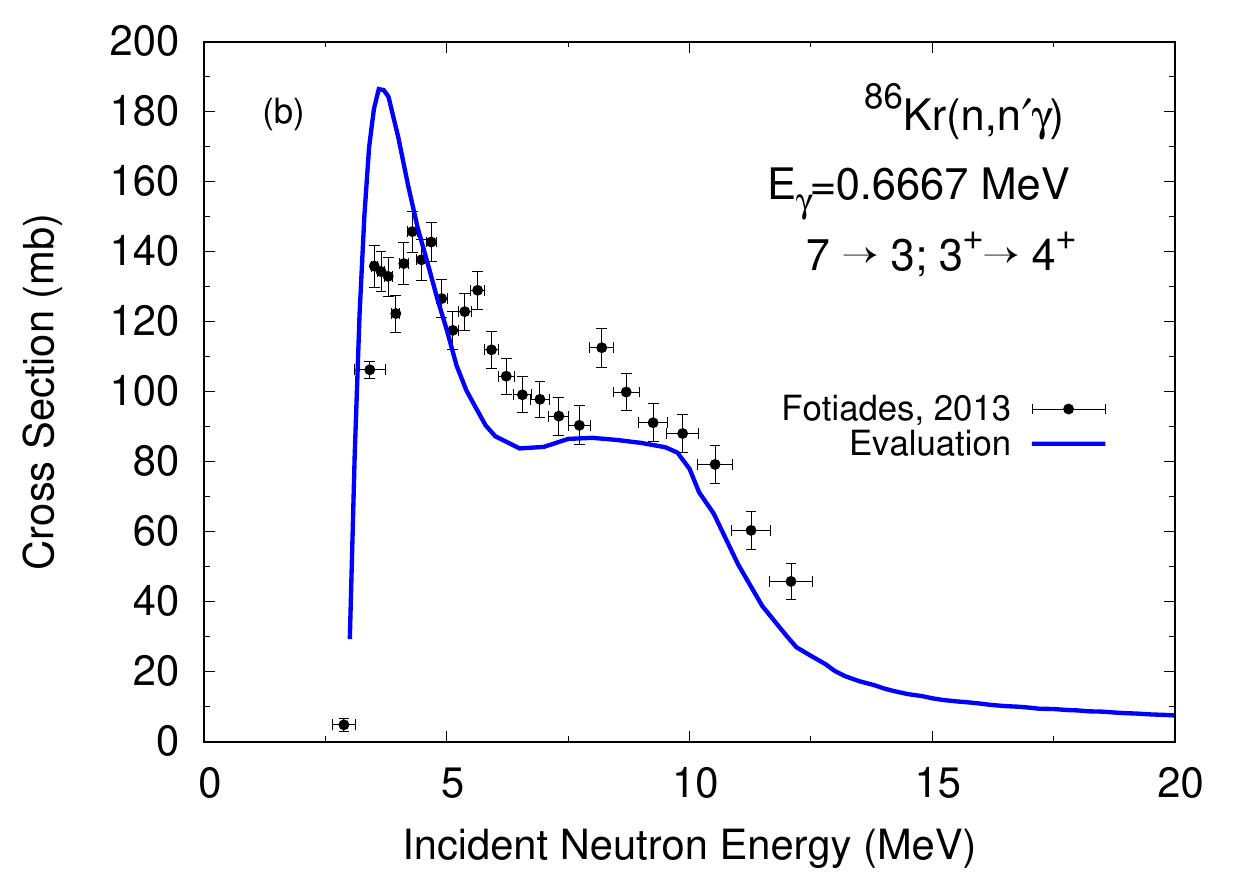}}
\subfloat{\includegraphics[scale=0.51, clip, trim = 9mm 7mm 4mm 2mm]{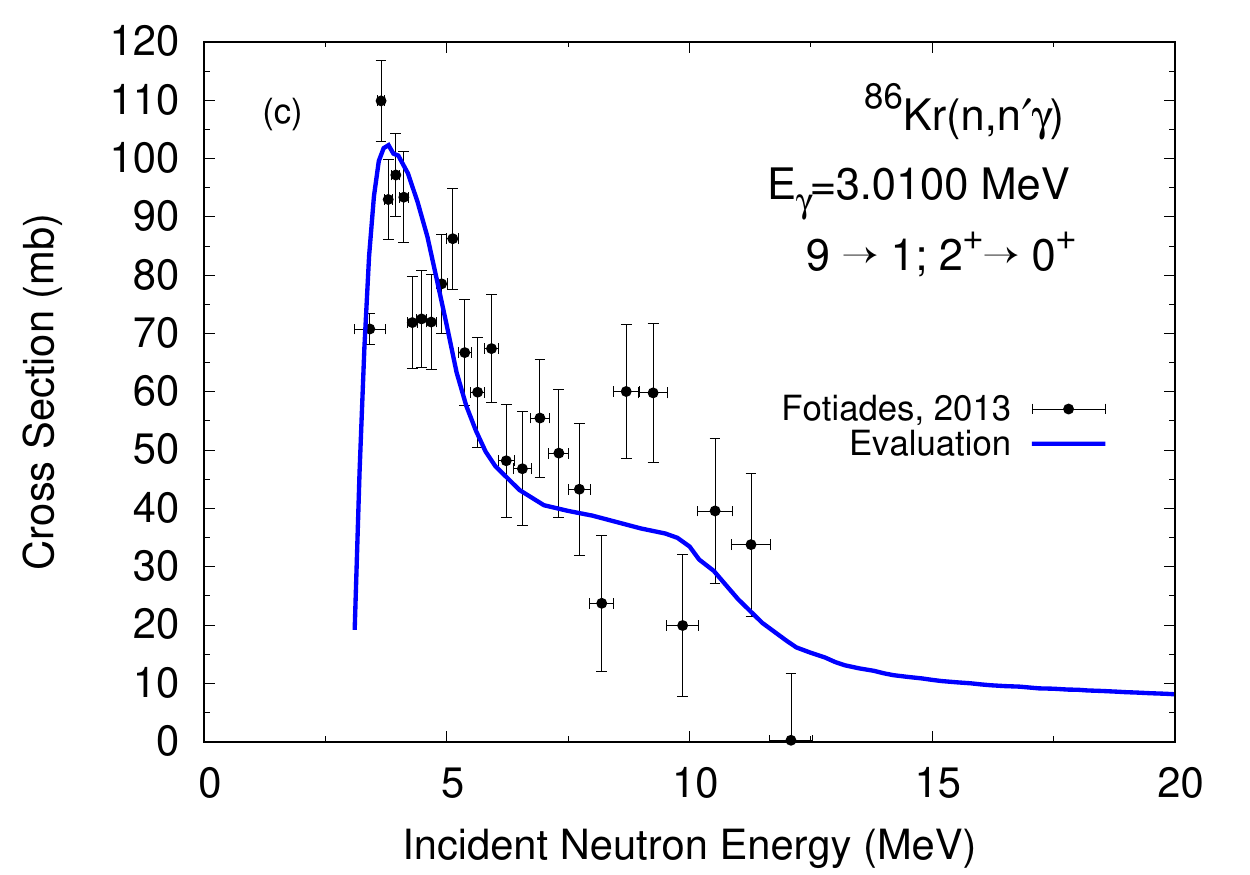}}
\\
\subfloat{\includegraphics[scale=0.51, clip, trim = 0mm 7mm 4mm 2mm]{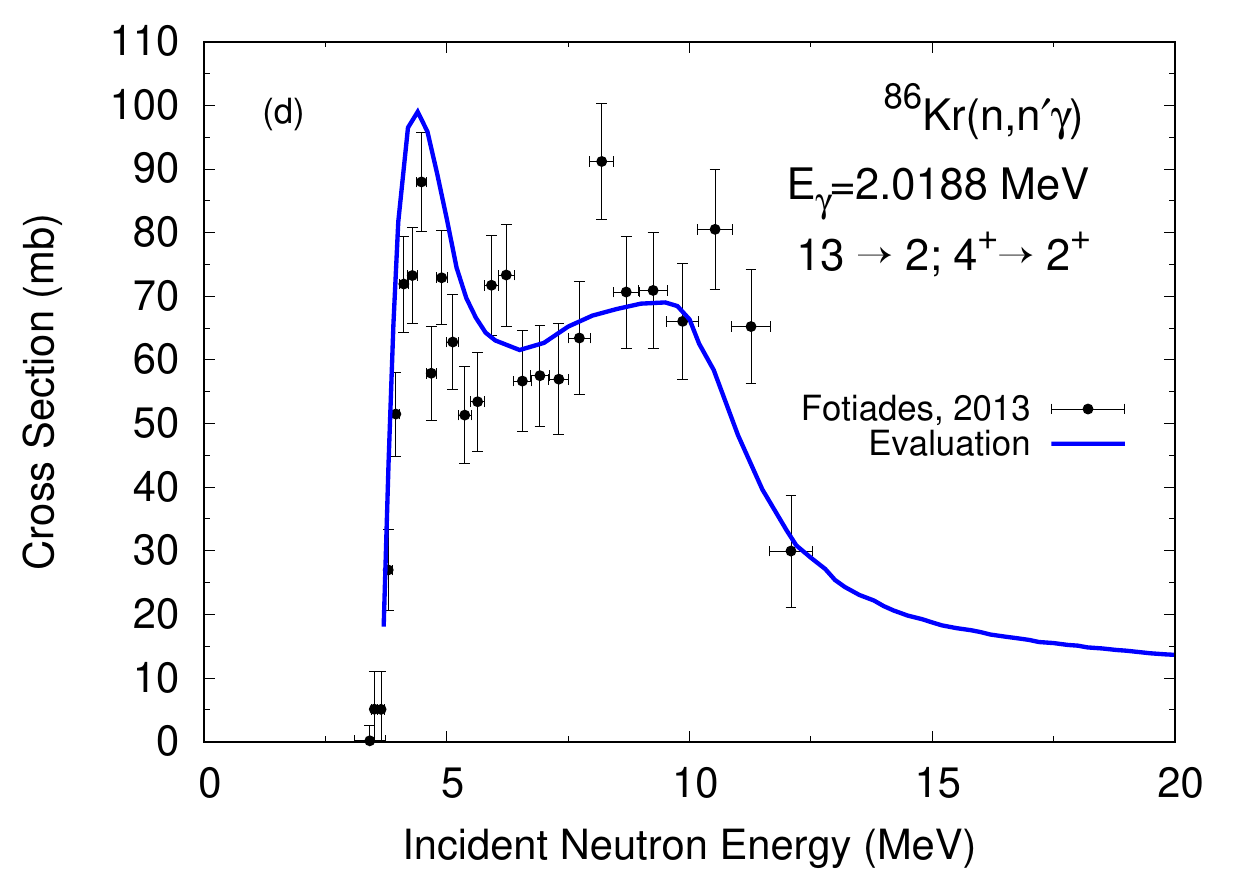}}
\subfloat{\includegraphics[scale=0.51, clip, trim = 9mm 7mm 4mm 2mm]{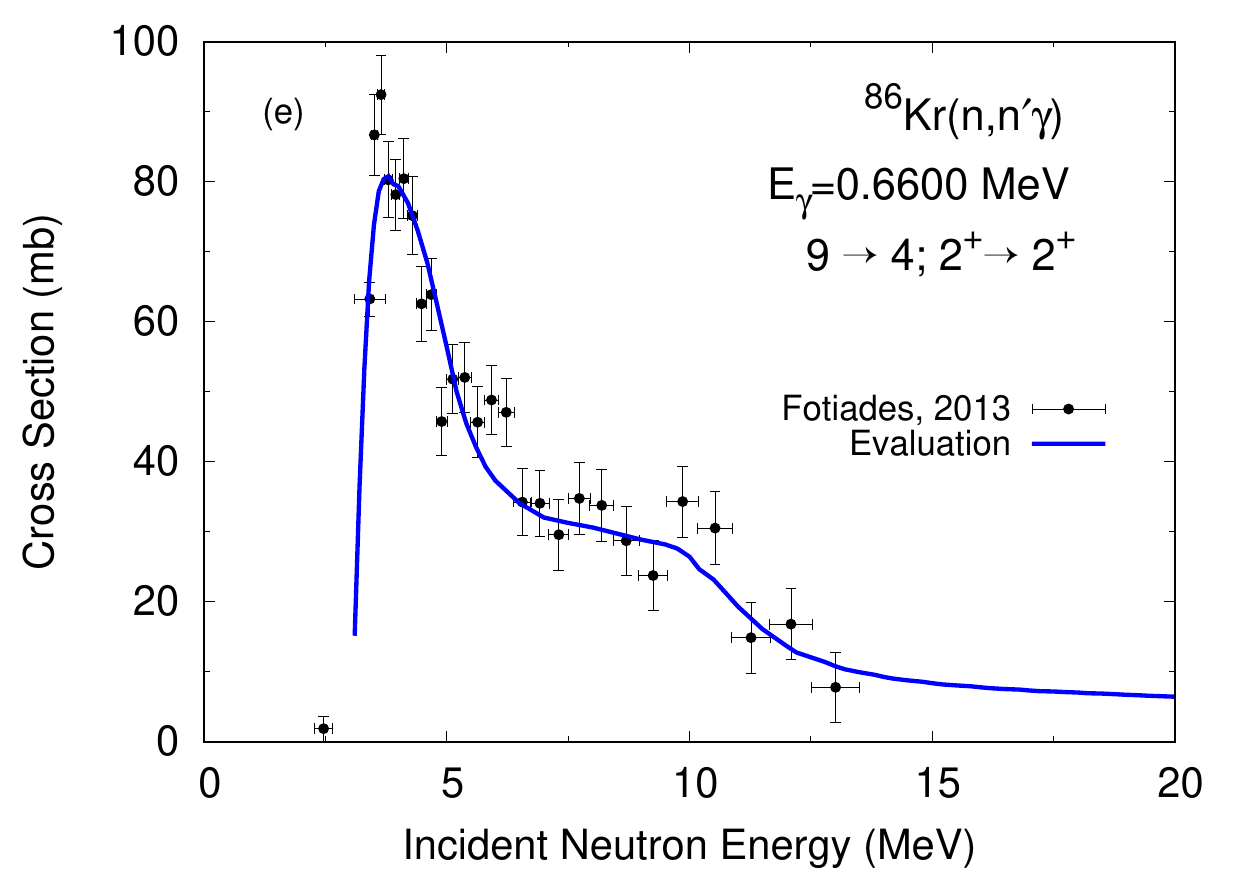}}
\subfloat{\includegraphics[scale=0.51, clip, trim = 9mm 7mm 4mm 2mm]{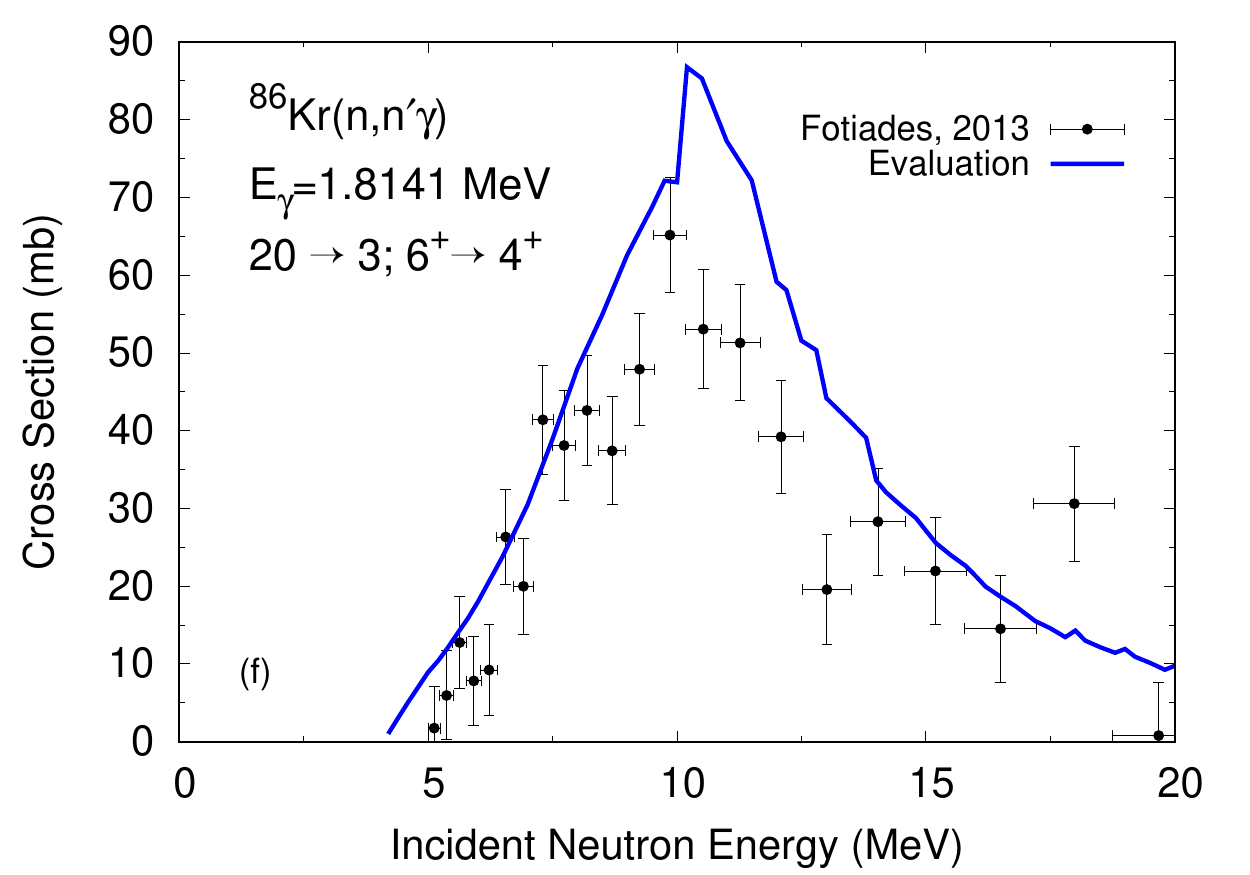}}
\\
\subfloat{\includegraphics[scale=0.51, clip, trim = 0mm 7mm 4mm 2mm]{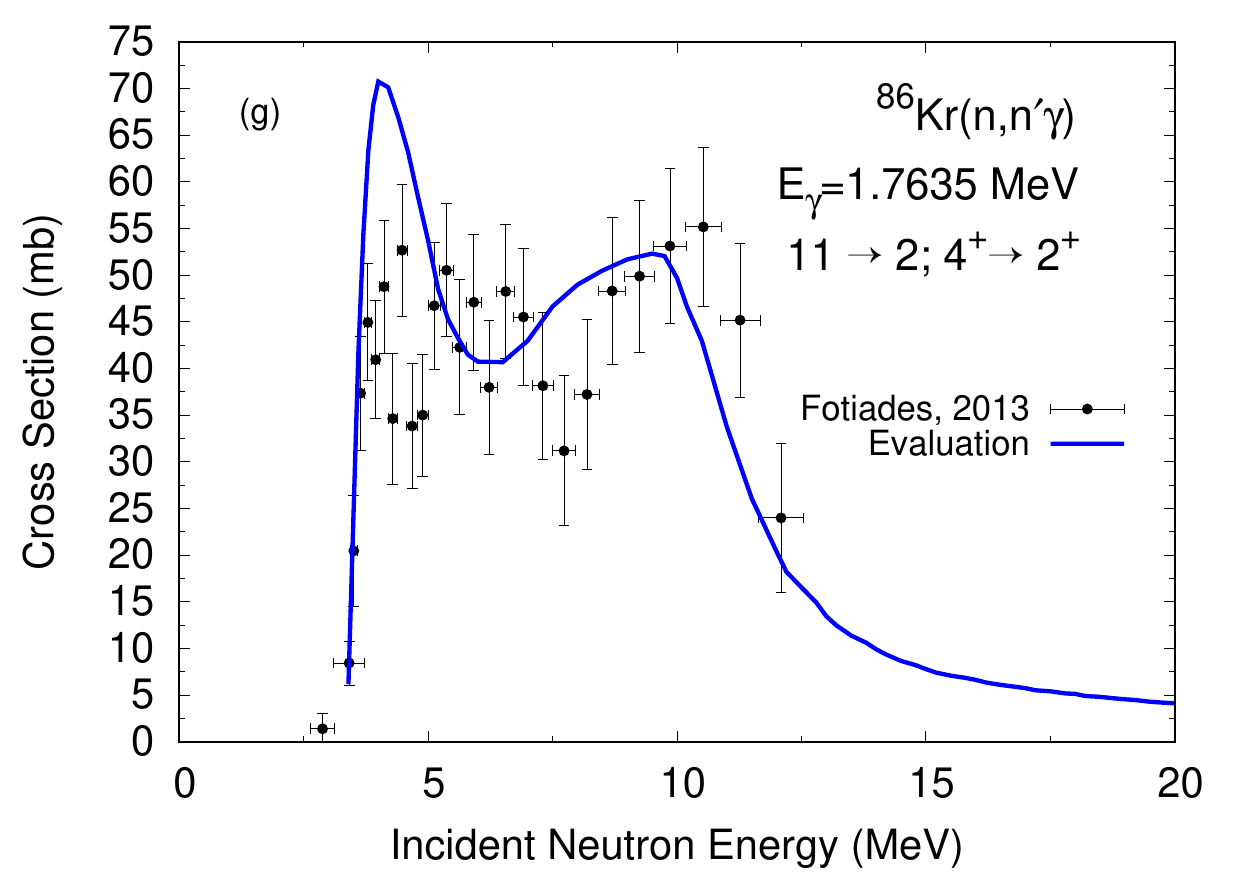}}
\subfloat{\includegraphics[scale=0.51, clip, trim = 9mm 7mm 4mm 2mm]{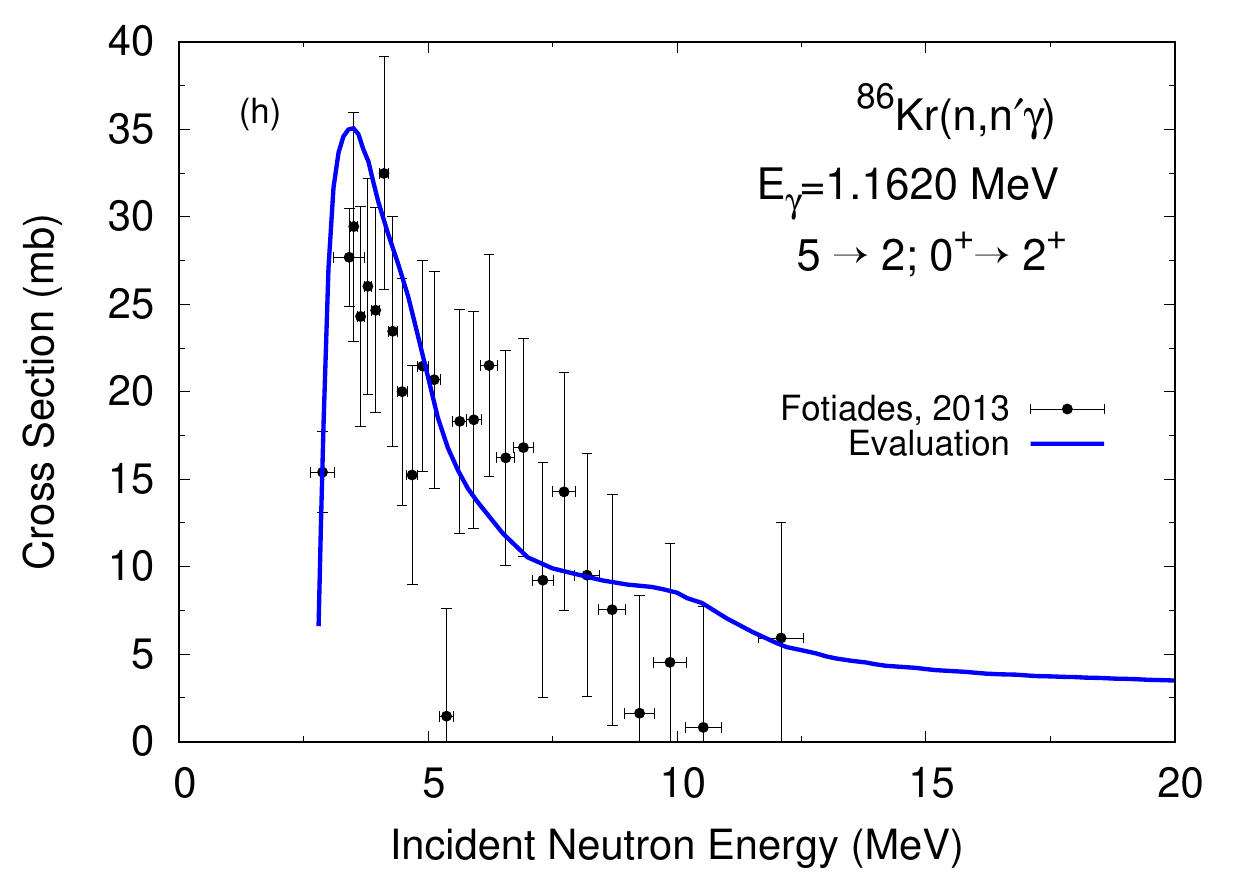}}
\subfloat{\includegraphics[scale=0.51, clip, trim = 9mm 7mm 4mm 2mm]{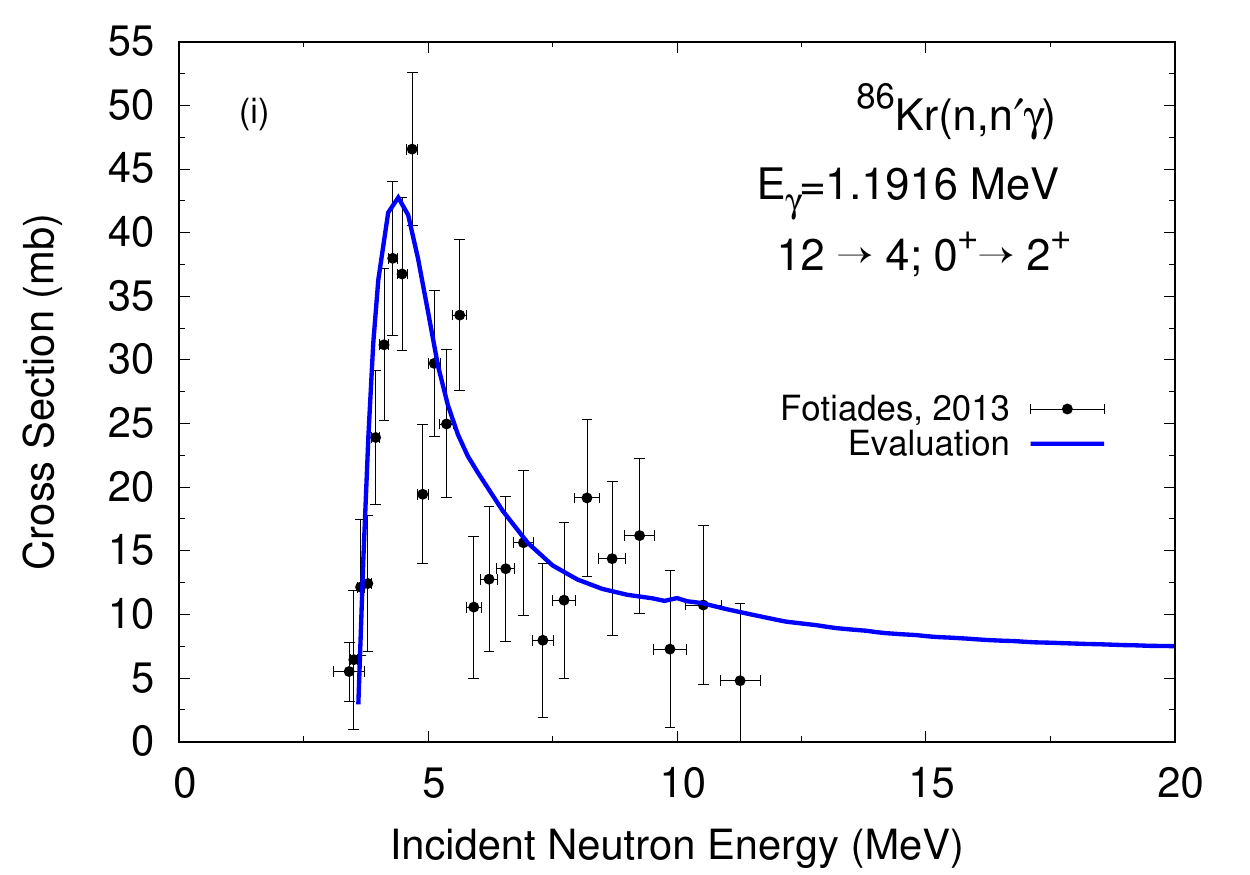}}
\\
\subfloat{\includegraphics[scale=0.51, clip, trim = 0mm 0mm 4mm 2mm]{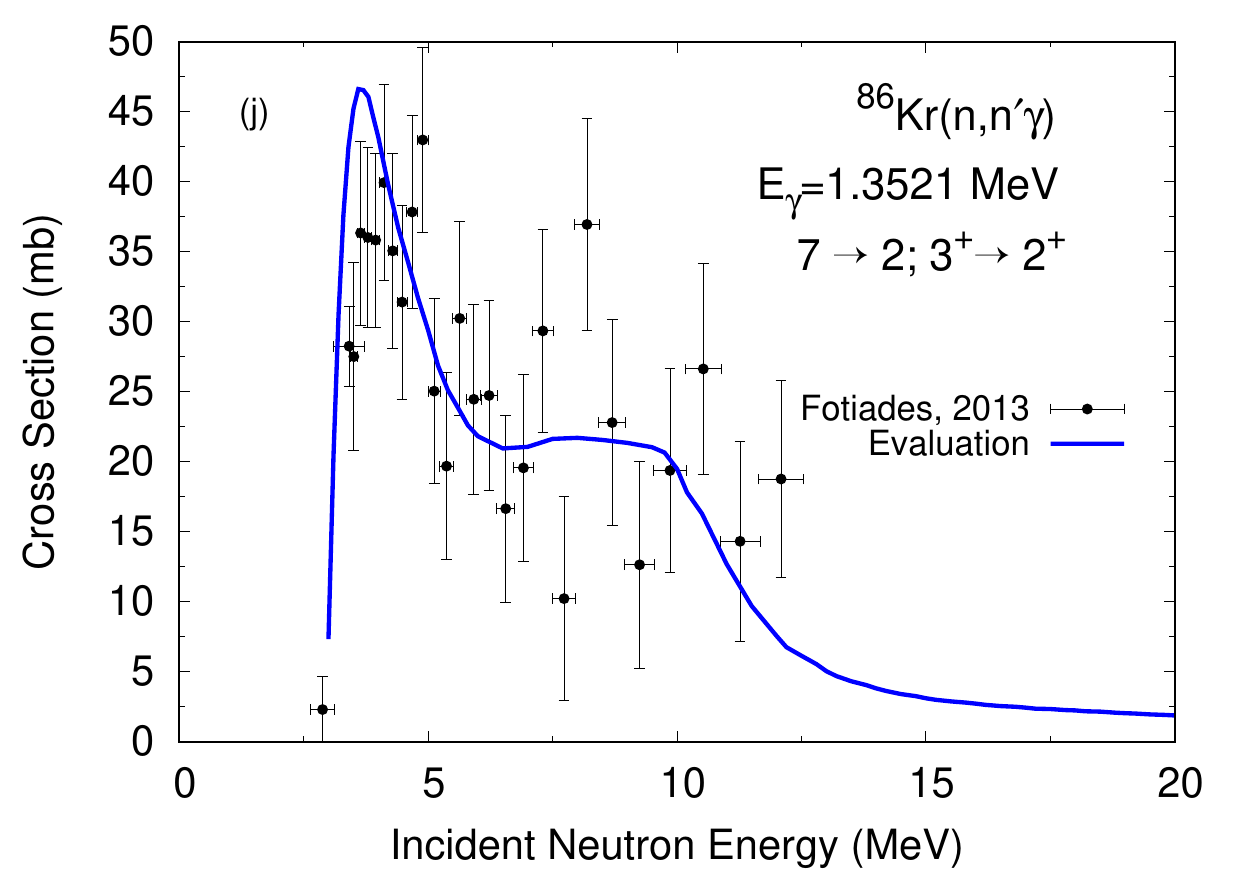}}
\subfloat{\includegraphics[scale=0.51, clip, trim = 9mm 0mm 4mm 2mm]{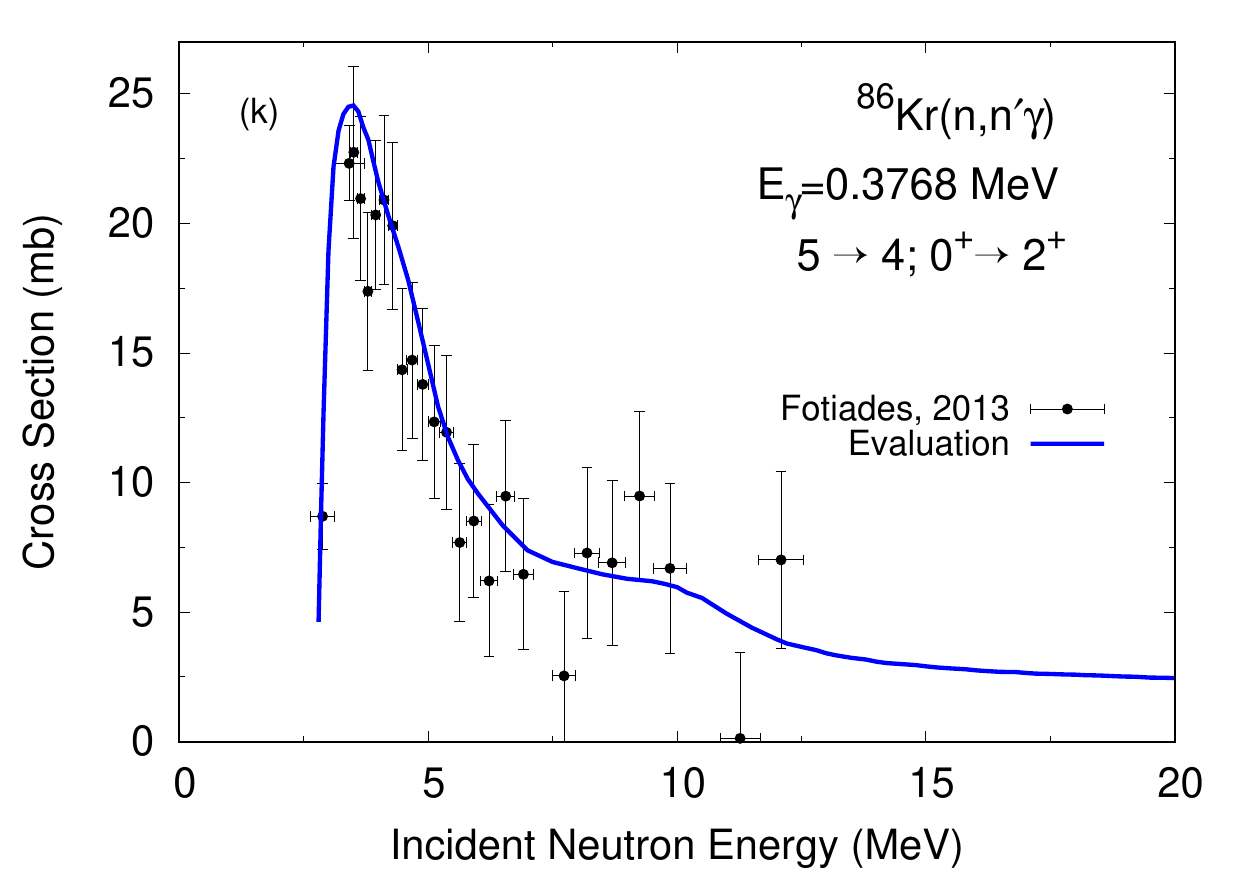}}
\subfloat{\includegraphics[scale=0.51, clip, trim = 9mm 0mm 4mm 2mm]{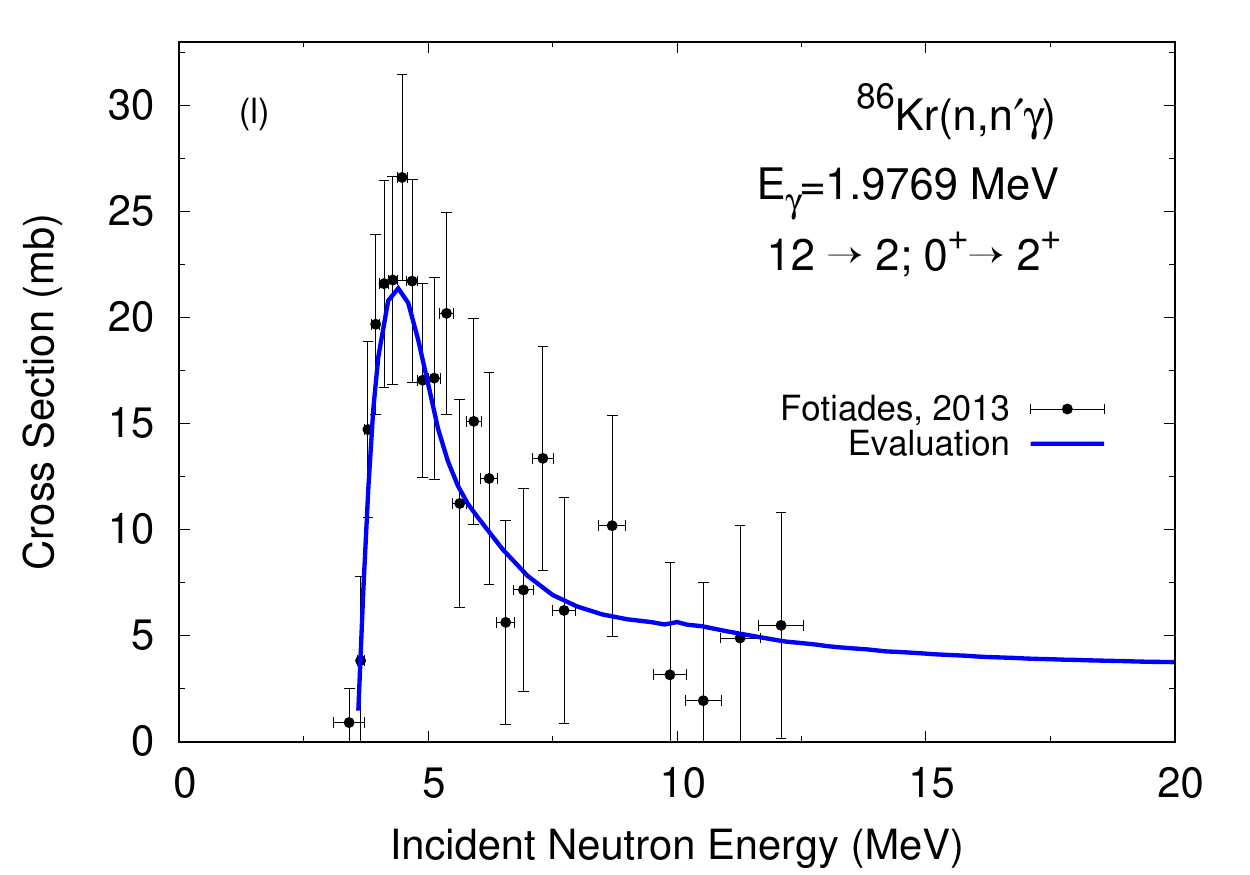}}
\caption{Inelastic gamma cross sections for discrete transitions between excited levels of \nuc{86}{Kr}. The indices for initial and final levels of the transitions, with their corresponding spin and parity, as well as the $\gamma$ energies, $E_\gamma$, are listed in the plots.
In this work we adopted the convention of representing the ground state with the index 1, the first excited state with the index 2, and so forth.
Our evaluation (blue curves) is compared to the experimental data measured by Fotiades \etal~\!\!\!\cite{Fotiades_2013} (black points). We do not show ENDF/B-VIII.0 here as inelastic gamma information was not present in that evaluation.}
\label{inel_transition_xsec}
\end{figure*}

In Fig.~\ref{first_inel_transition_xsec} we display the comparison
between the data and our evaluation for the first inelastic transition while all the other ones are shown in Fig.~\ref{inel_transition_xsec}. In this case there is no comparison with
ENDF/B-VIII.0~\cite{Brown:2018} because any $\gamma$-ray production information is currently provided in the library. As for the case of the total cross section,
the remarkable agreement between our calculation and the data for the $2^+ \rightarrow 0^+$ transition is not surprising since it is the result of a fit of the optical potential for energies up
to $12$ MeV and the pre-equilibrium model for higher energies. On the contrary, it is interesting to notice how the fit of the optical potential and the pre-equilibrium model were able
to produce results for all the other transitions that are overall in good agreement with the data. In particular, it is worth noting how the data for the $4^+ \rightarrow 2^+$ transition,
showed in the top left panel of Fig.~\ref{inel_transition_xsec}, is remarkably well reproduced.

\subsubsection{Neutron Production Cross Section}
\begin{figure}[t]
\includegraphics[width=1\columnwidth]{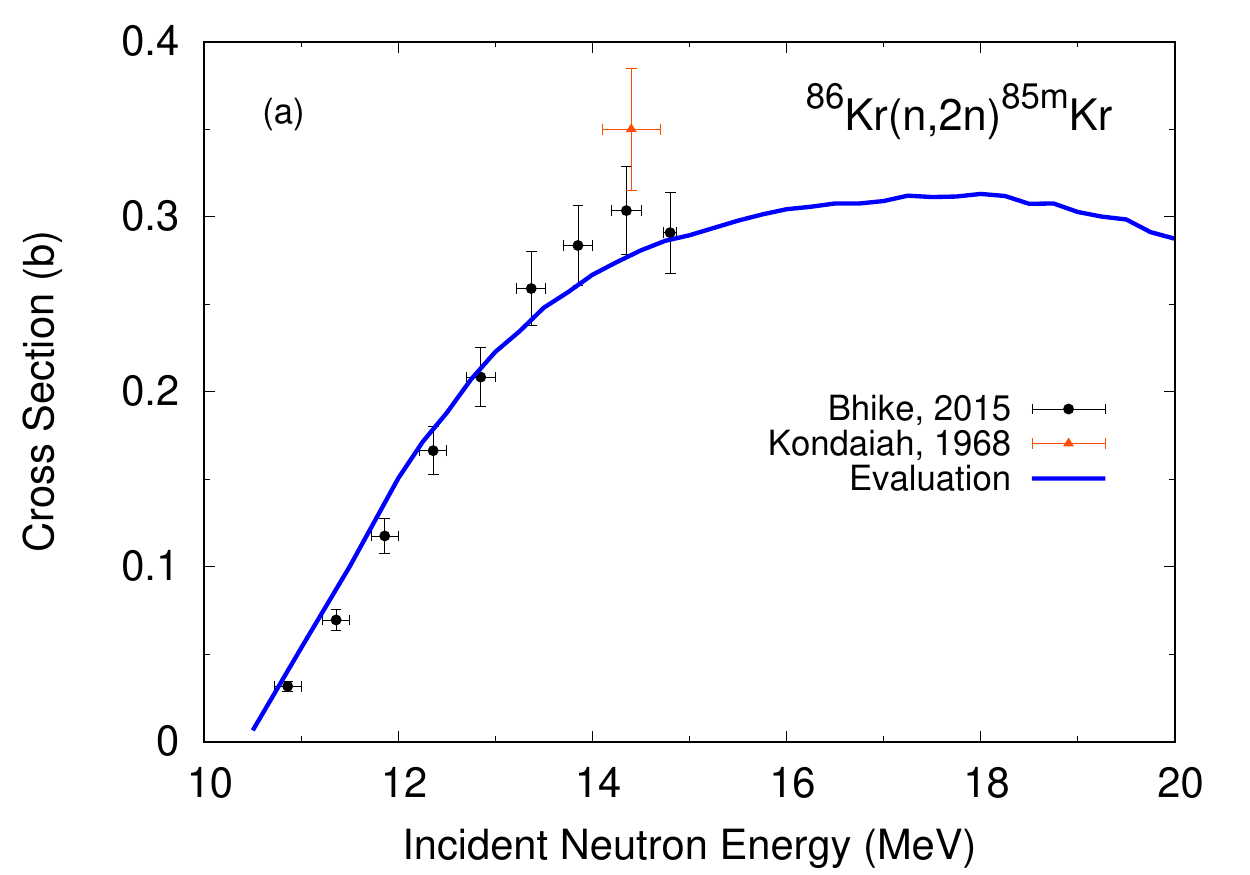}
\includegraphics[width=1\columnwidth]{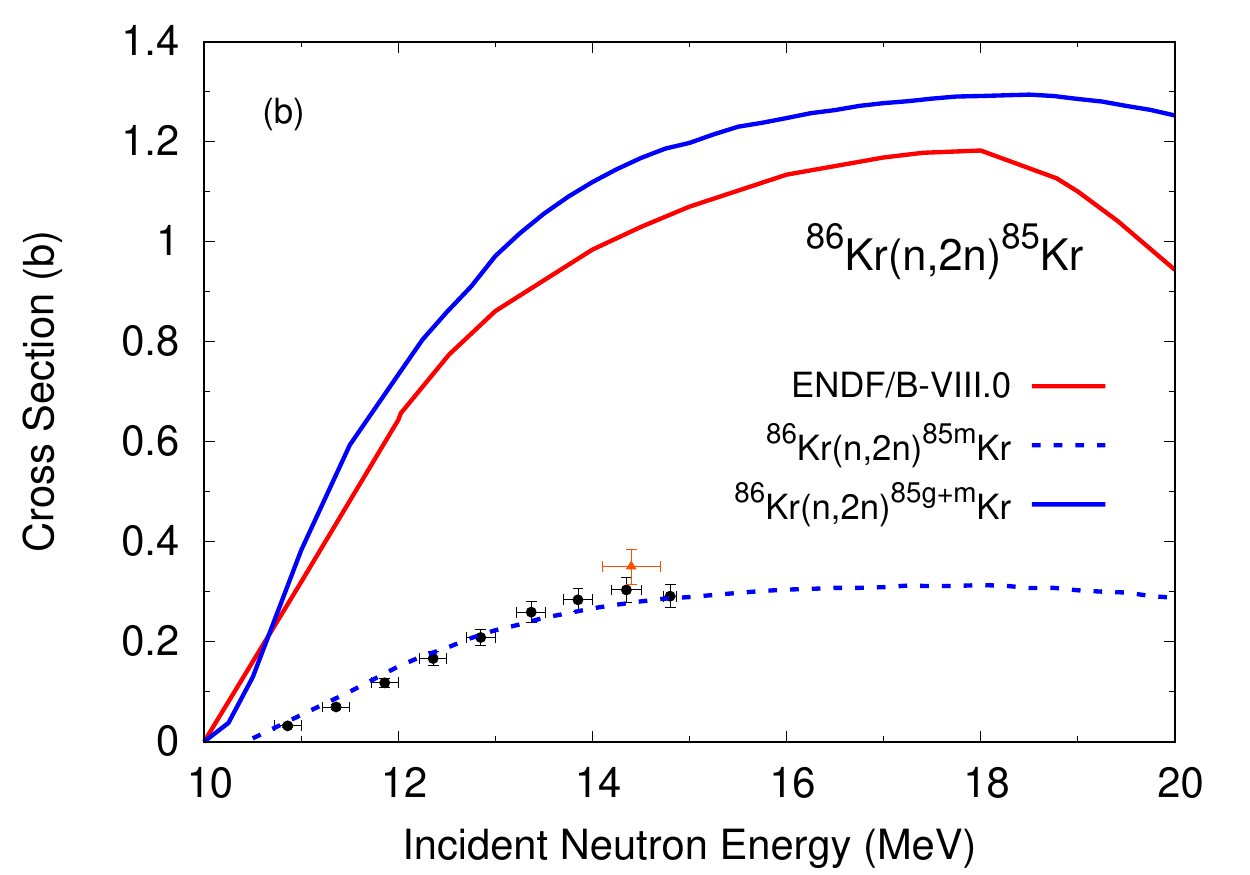}
\caption{Panel (a): evaluation of the $(\mathrm{n},2\mathrm{n})$ isomer cross section. Two sets of experimental data~\cite{Kondaiah_1968,Bhike_2015} are presented along
with the evaluated cross section. Panel (b): Comparison of the evaluated total $(\mathrm{n},2\mathrm{n})$  cross section with ENDF/B-VIII.0. No experimental data are available for this observable. For reference, the isomeric cross section and data from panel (a) are also shown.
}
\label{n_2n_xsec}
\end{figure}

The $^{86}$Kr$(\mathrm{n},2\mathrm{n}) ^{85m}$Kr reaction is important for both the NIF diagnostics and for understanding the $^{85}$Kr branching point in the s-process.
The first excited state of $^{85}$Kr is an isomer, and the decay of this isomeric state also provides a convenient method to measure the total $(\mathrm{n},2\mathrm{n})$ reaction rate in RAGS detector, providing information on the primary fusion neutrons.
Two sets of experimental data are available for the $^{86}$Kr$(\mathrm{n},2\mathrm{n}) ^{85m}$Kr isomer reaction.
One set was measured at the Triangle Universities Nuclear Laboratory (TUNL) by Bhike \etal~\cite{Bhike_2015}, in the energy range between $10$ MeV and $15$ MeV.
A single measurement of the $^{86}$Kr$(\mathrm{n},2\mathrm{n}) ^{85m}$Kr cross section was taken by Kondaiah \etal~\!\!\!\cite{Kondaiah_1968}, at the Georgia Tech accelerator
at $14.4$ MeV.
Fig.~\ref{n_2n_xsec} presents the experimental data for the isomer reaction, along with the evaluated cross section, which was calculated with EMPIRE~\cite{empire} tuning the exciton
model to reproduce the shape of the data of Bhike \etal We favored that experiment as it is much more recent and complete than the one from Kondaiah \etal

\subsubsection{Radiative Capture Cross Section}
\begin{figure}[t]
\includegraphics[width=1\columnwidth]{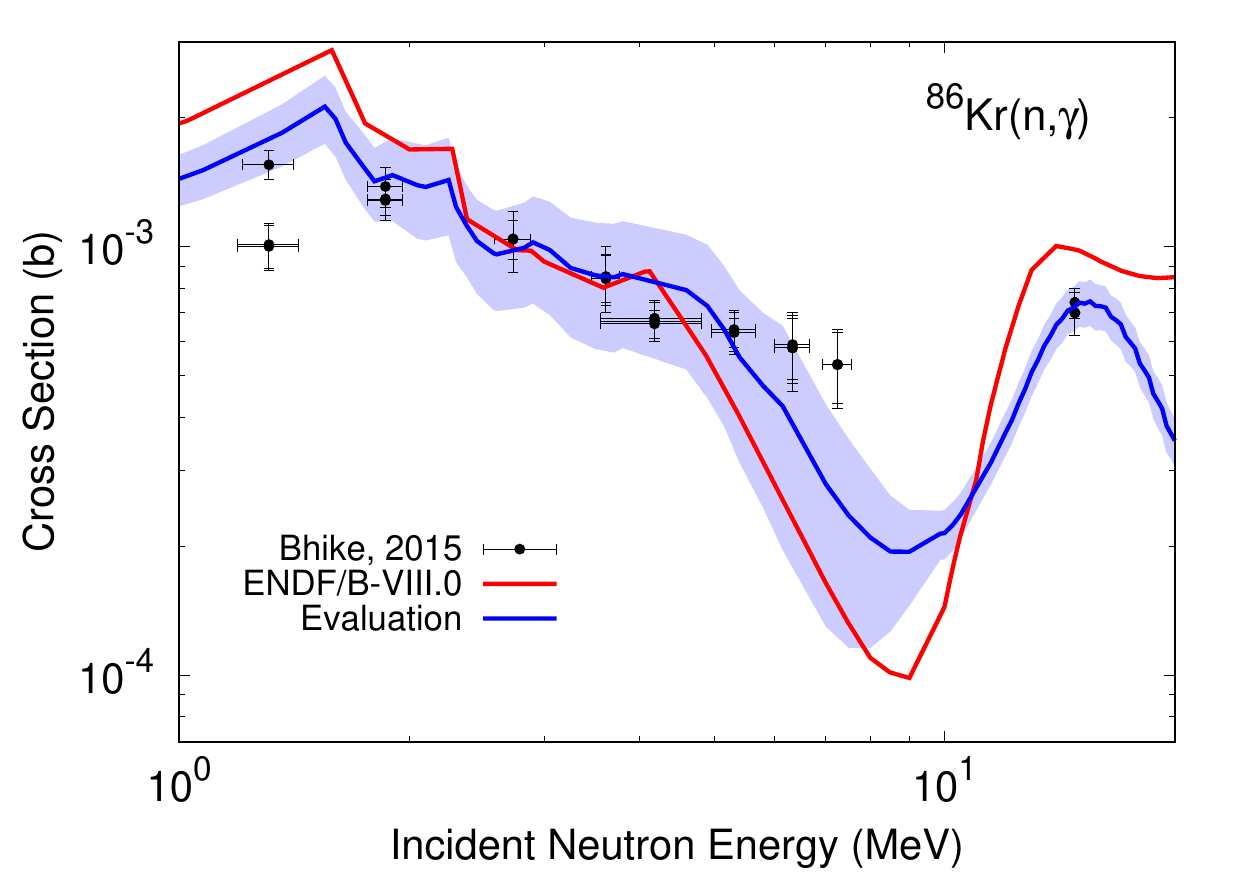}
\caption{Evaluation of the $(\mathrm{n},\gamma)$ cross section compared with ENDF/B-VIII.0~\cite{Brown:2018} and the experimental data~\cite{Bhike_2015}.
The light-blue band represents the uncertainty band of the cross section and it is obtained form the covariance matrix.}
\label{capture_xsec}
\end{figure}
The $^{86}$Kr$(\mathrm{n},\gamma ) ^{87}$Kr reaction is important for studying the low energy component of the NIF flux, and for calculating the abundance of $^{87}$Kr in AGB stars.
Bhike \etal~\cite{Bhike_2015} measured the radiative capture cross section in the energy region between $0.5$ MeV and $15$ MeV in 2015.
This set of experimental data is presented in Fig.~\ref{capture_xsec}, along with the calculated and evaluated cross section that is also compared with ENDF/B-VIII.0~\cite{Brown:2018}.
We also display in light blue the uncertainty band obtained from the covariance matrix.

\subsubsection{Partial Alpha Production Cross Section}
\begin{figure}[hbpt]
\includegraphics[width=1\columnwidth]{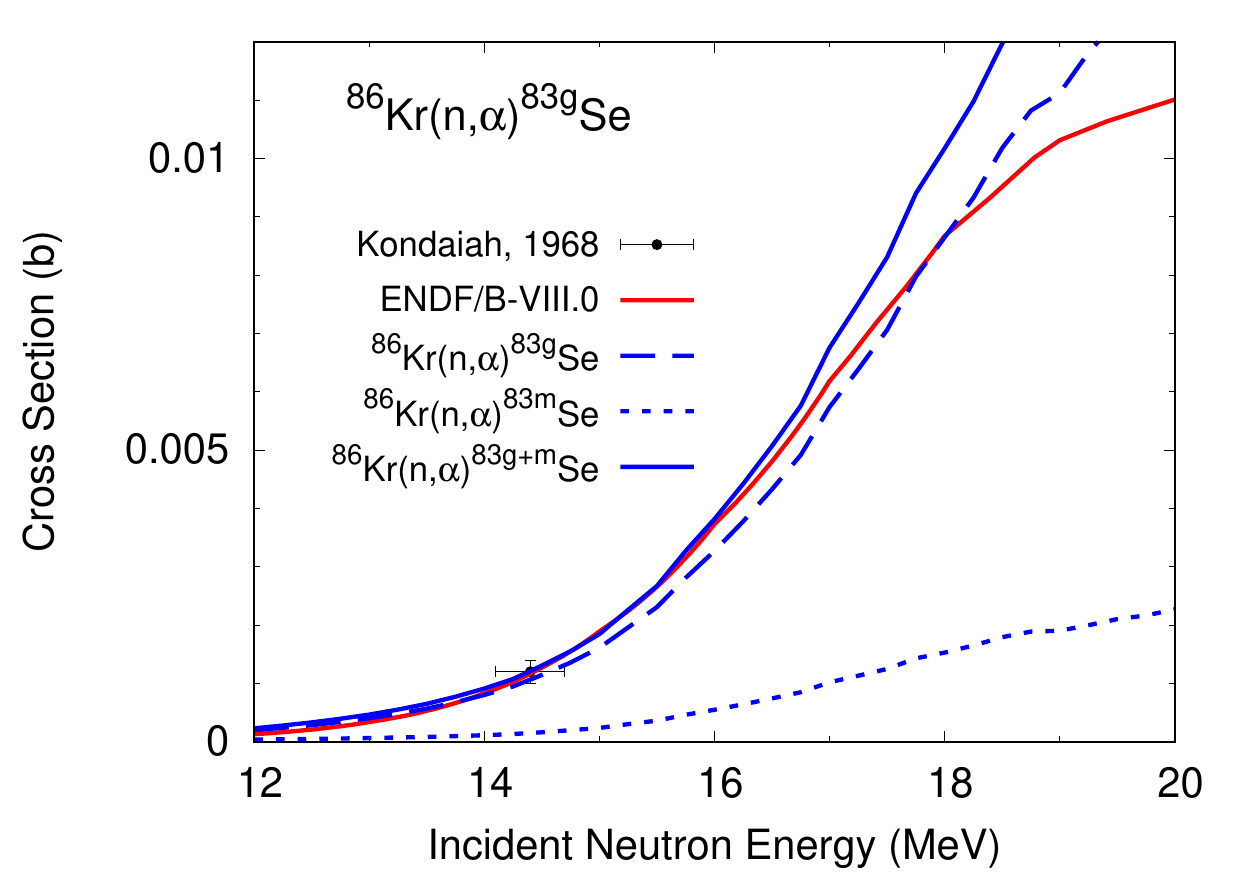}
\caption{Evaluation of the $(\mathrm{n},\alpha)$ cross section compared with the only available data taken from Ref.~\cite{Kondaiah_1968}, which corresponds to the residual nucleus \nuc{83}{Se} in the ground state. Since no other information is
available, we tune the alpha production mechanism to reproduce this data.}
\label{alpha_xsec}
\end{figure}
For this case we only have one data point from Kondaiah \etal~\!\!\!\cite{Kondaiah_1968} for the $^{86}$Kr(n,$\alpha$)$^{83\mathrm{g}}$Se process.
Since no other information is currently available we tuned the $\alpha$ production mechanism to reproduce this data.
Our evaluated cross section for this process is displayed in Fig.~\ref{alpha_xsec} along with the cross section for the isomeric state and their sum.
It is worth notice that there is a significant difference between our results and ENDF/B-VIII.0~\cite{Brown:2018}, because the latter does not include any
isomeric cross section and the result displayed in red in Fig.~\ref{alpha_xsec} represents the total cross section, that was fitted to reproduce the only available point
for the $^{86}$Kr(n,$\alpha$)$^{83\mathrm{g}}$Se process.

\subsubsection{Neutron Emission Spectra}
   \label{subSec:de}
In general, there are several measurements of neutron energy spectra concentrated around 14 MeV for many nuclei.
At these incident energies, the high outgoing-energy end of the spectra is dominated by the elastic peak and direct transitions to the collective levels that are well modeled by CC and DWBA calculations. The middle part of the spectrum is governed by the pre-equilibrium emission,  while the low energy peak is the evaporation peak described by the Hauser-Feshbach model. Proper description of the neutron spectra is thus a very good overall test of the quality of the reaction modeling used in the evaluation.


Energy-angle correlated cross sections add another dimension to the energy spectra, associated with the scattering angle.  These arguably correspond to an even more direct kind of measurement since energy spectra are usually obtained by integrating double-differential ones which unavoidably involves some approximations related to interpolation and extrapolation of angular distributions.
However, employed models in EMPIRE~\cite{empire} rely on empirical Kalbach parameterization of double-differential cross sections \cite{Kalbach:81}.
Therefore, comparison of an evaluation with  double-differential cross sections provides for an additional stringent test of the modeling employed in the evaluation procedure. 

\begin{figure*}[hbpt]
\includegraphics[scale=0.70,clip,trim =   0mm 7mm 0mm 0mm]{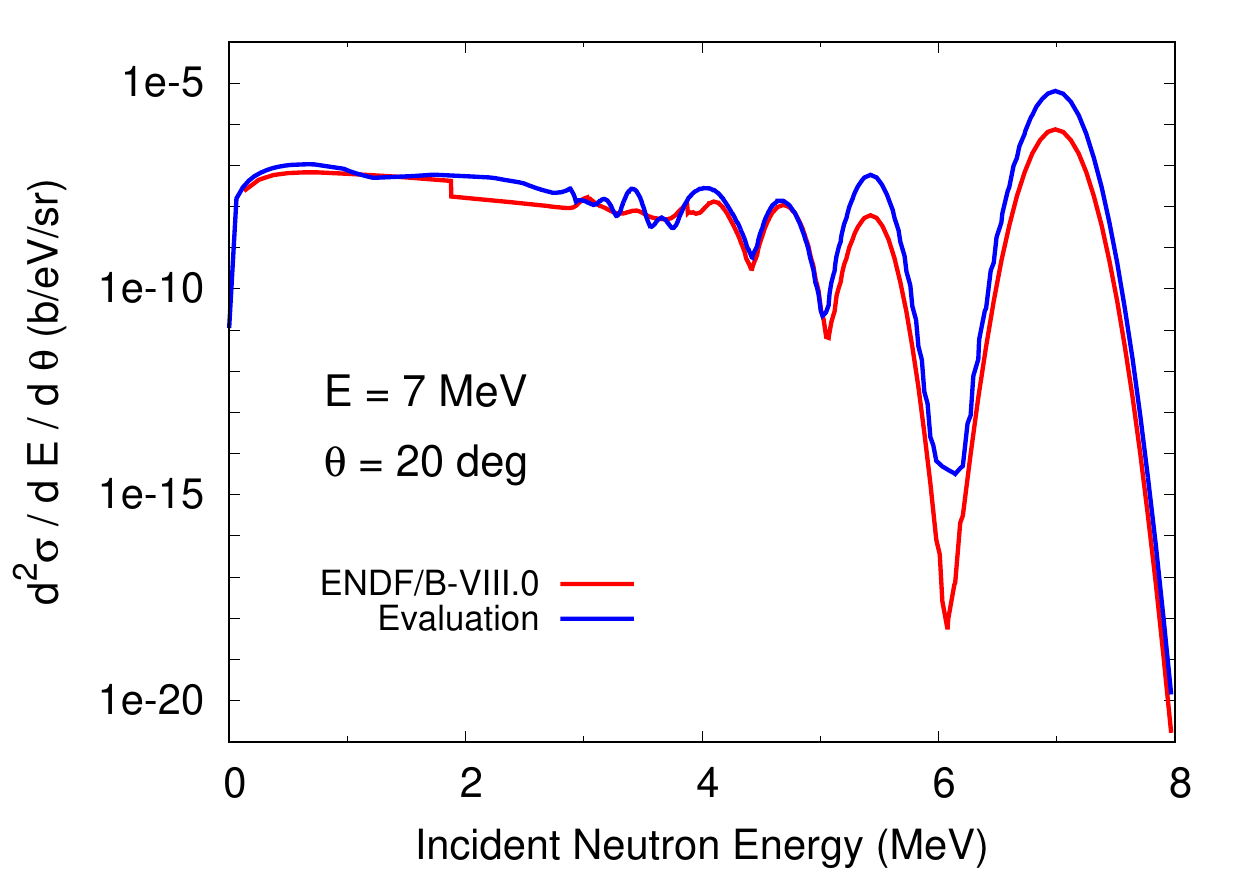}
\includegraphics[scale=0.70,clip,trim = 10mm 7mm 0mm 0mm]{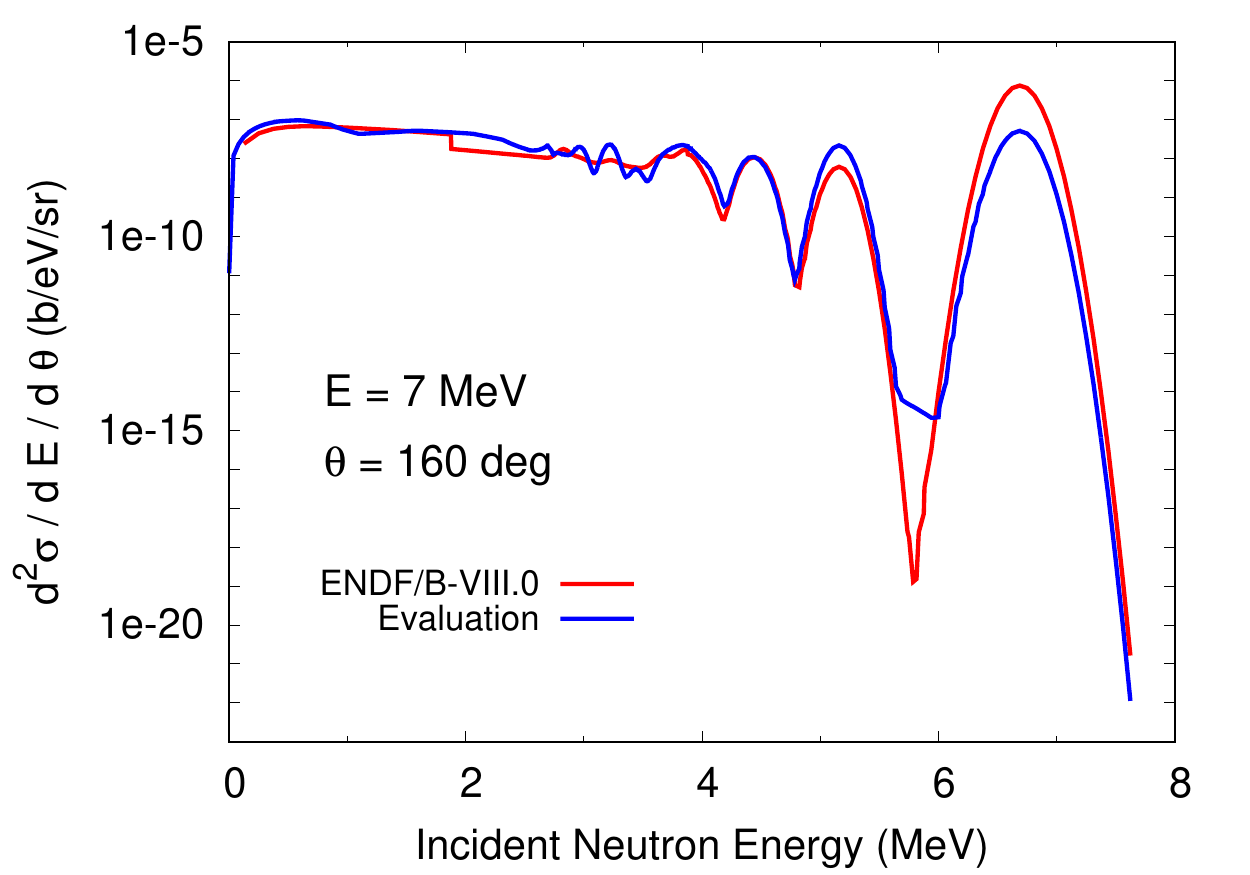}
\\
\includegraphics[scale=0.70,clip,trim =   0mm 7mm 0mm 0mm]{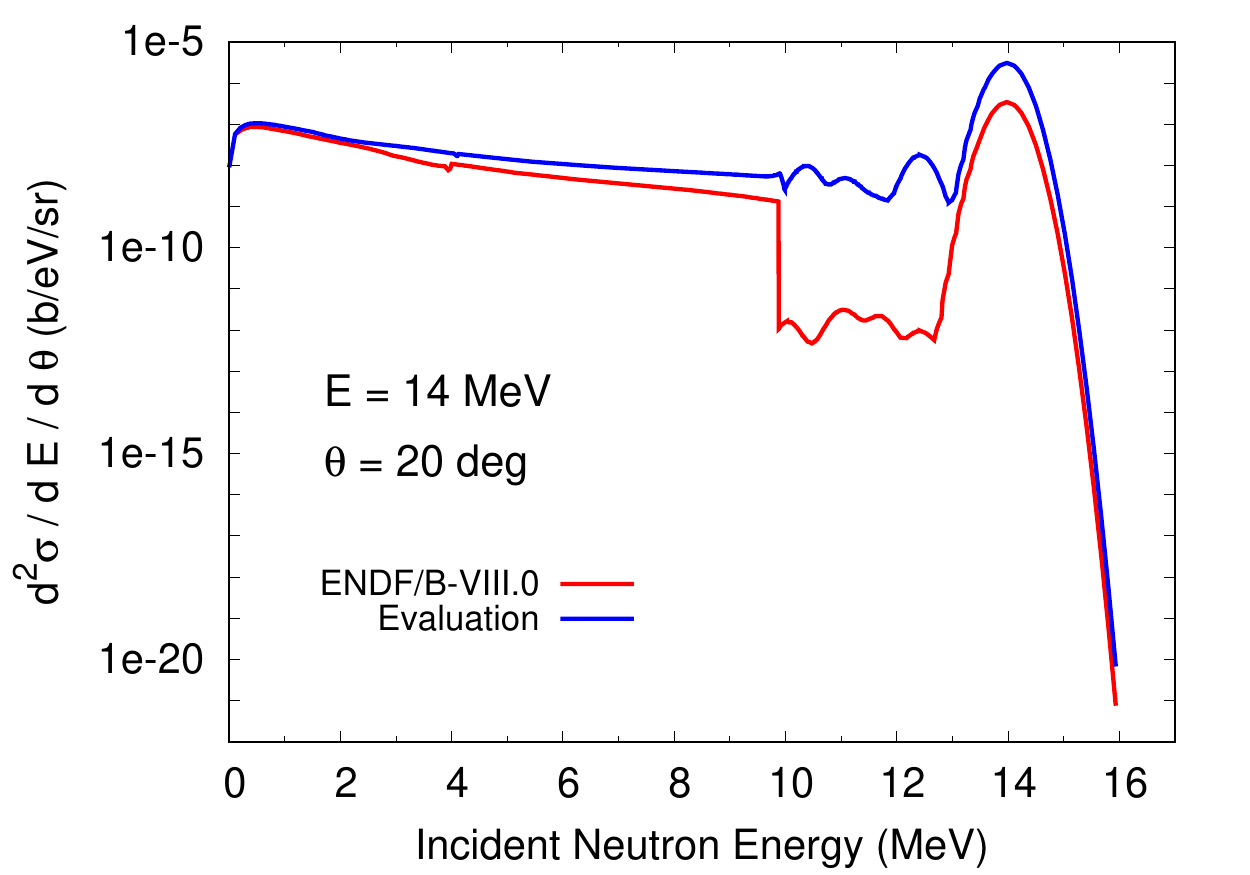}
\includegraphics[scale=0.70,clip,trim = 10mm 7mm 0mm 0mm]{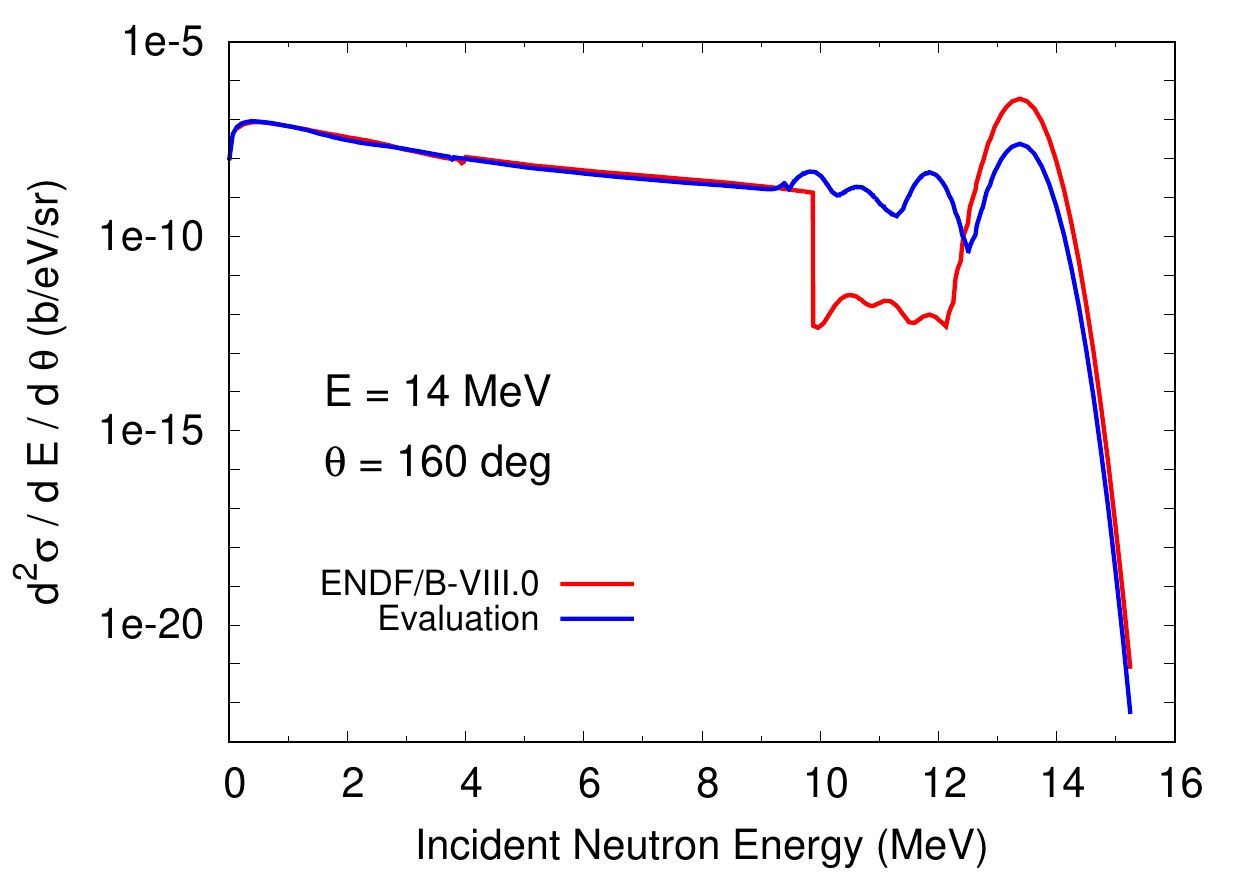}
\\
\includegraphics[scale=0.70,clip,trim =   0mm 0mm 0mm 0mm]{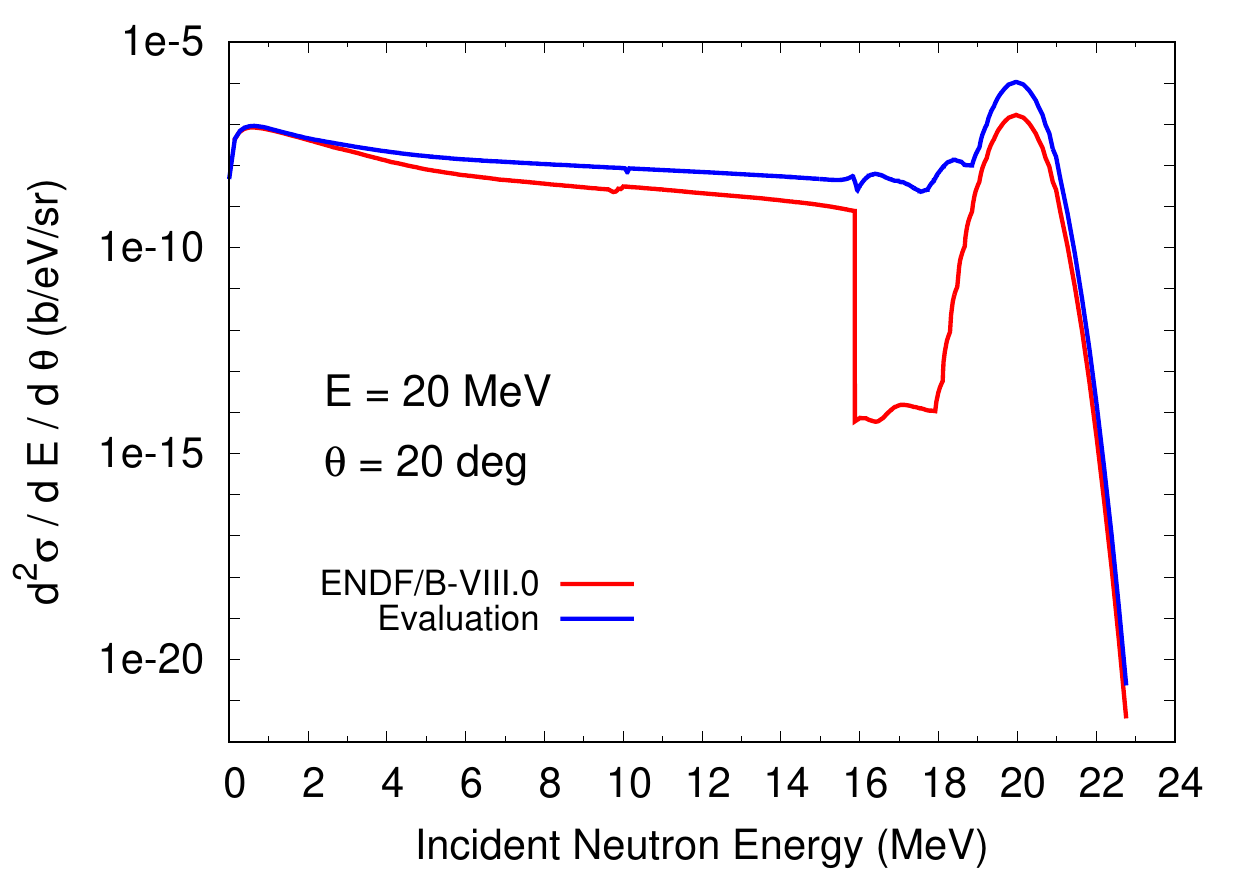}
\includegraphics[scale=0.70,clip,trim = 10mm 0mm 0mm 0mm]{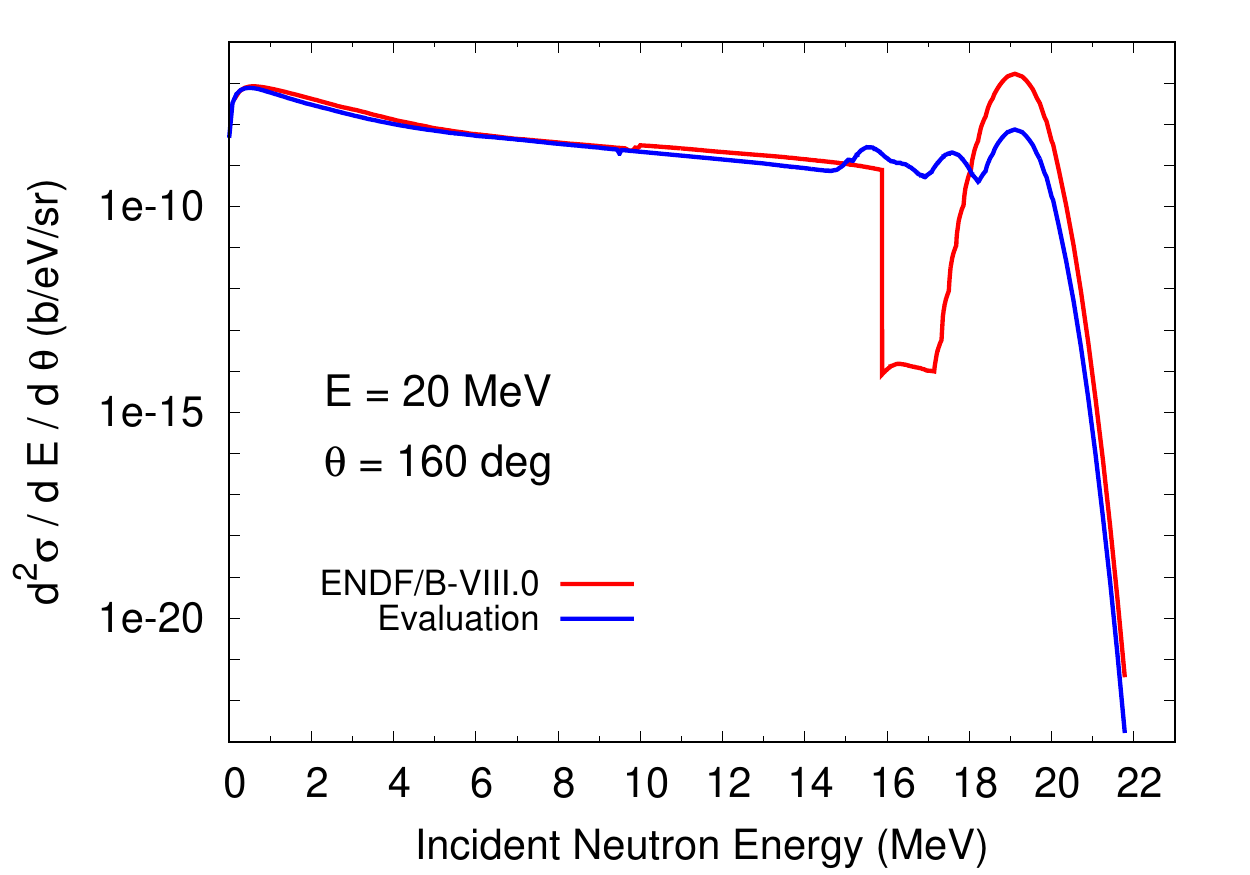}
\caption{Double differential cross sections computed at different incident energies and scattering angles compared with ENDF/B-VIII.0~\cite{Brown:2018}.}
\label{dd_xsec}
\end{figure*}

Unfortunately, in the specific case of \nuc{86}{Kr} there are no spectra data available to guide our model parametrization. However, a direct comparison of our evaluation with ENDF/B-VIII.0~\cite{Brown:2018}, as seen in Fig.~\ref{dd_xsec} for different scattering angles and values of incident energy, shows that our careful description of inelastic couplings and pre-equilibrium provides a much more realistic spectra, without the gaps seen in ENDF/B-VIII.0. Therefore, our evaluation represents a major improvement in the the description of the inelastic process.


%

%
%
%
%

\section{Conclusions}\label{Sec:conclusions}
A new $^{86}$Kr evaluation was performed based on a new resolved resonance region evaluation and several sets of experimental data.
The calculation of the neutron-induced reactions was performed using an optical model potential properly fitted to reproduce the experimental data for the total cross section and
the main inelastic gamma transition $2^+ \rightarrow 0^+$ for energies up to 12 MeV. Due to poor statistics and the consequent large fluctuations in the data for the total cross section,
the Koning-Delaroche optical model potential was adopted at higher energies along with the exciton model that was properly tuned to reproduce the $(\mathrm{n},2\mathrm{n})$
experimental data.
Several level $J^{\pi}$ assignments and branching ratios were also fixed in the $^{86}$Kr structure, leading to an overall evaluation that fits the experimental data.
The accurate evaluation of the neutron production and radiative capture reactions will improve the modeling and calculations for the RAGS detector at NIF and understanding of the
$^{85}$Kr branching point in the s-process.
The new inelastic partial gamma evaluation provides insight into the nuclear structure of $^{86}$Kr and may be used in future structure evaluations.

The evaluation presented here is a major improvement when compared to the  previous ENDF/B-VIII.0~\cite{Brown:2018} file as it not only reproduces new measured data in detail but also corrects issues with thermal capture and total cross sections in the RRR, as well as provides data-constrained inelastic $\gamma$ production and isomeric cross sections. Inelastic spectra are also much more realistic even though the lack of data prevented us to describe it in more detail.
This new evaluation of the neutron-induced reactions on $^{86}$Kr provides for the first time recommended values for $\gamma$-ray production, extremely important for the
NIF diagnostics, and isomeric cross section production, and it has been submitted to be considered for inclusion in the next release of the ENDF/B library.

\acknowledgments{
 Work at Brookhaven National Laboratory was sponsored by the Office of Nuclear Physics, Office of Science of the U.S. Department of Energy under Contract No. DE-SC0012704 with Brookhaven Science Associates, LLC. 
 This project was supported in part by the U.S. Department of Energy, Office of Science,
Office of Workforce Development for Teachers and Scientists (WDTS) under the
Science Undergraduate Laboratory Internships Program (SULI).
We would like to thank Dr. Balraj Singh, Dr. Alexandru Negret, and Dr. Elizabeth McCutchan for elucidative discussions on nuclear structure and level-spin assignments.

\bibliography{krypton_paper} 

\end{document}